\documentclass[prb,aps,twocolumn,showpacs,showkeys,amsmath,amssymb,floatfix,superscriptaddress]{revtex4-1}
\usepackage{booktabs,graphicx,verbatim,ulem}
\usepackage[colorlinks=true,citecolor=blue,filecolor=blue,linkcolor=blue,urlcolor=blue]{hyperref}
\usepackage{breakurl}
\newcommand{\1}[1]{\, \mathrm{#1}} 
\newcommand{\n}[1]{\mathrm{#1}}    

\newcommand{\columbia}{\affiliation{Physics Department, Columbia University, New York, NY 10027, USA}}
\newcommand{\bologna}{\affiliation{University of Bologna and INFN-Bologna, Bologna, Italy}}

\begin{document}

\title{Measurement of the Scintillation Yield of Low-Energy Electrons in Liquid Xenon}

\author{E.~Aprile}\columbia
\author{R.~Budnik}\columbia
\author{B.~Choi}\columbia
\author{H.~A.~Contreras}\columbia
\author{K.-L.~Giboni}\columbia
\author{L.~W.~Goetzke}\columbia
\author{J.~E.~Koglin}\altaffiliation[Present Address: ]{SLAC National Accelerator Laboratory, Menlo Park, CA 94025, USA}\columbia
\author{R.~F.~Lang}\altaffiliation[Present Address: ]{Physics Department, Purdue University, West Lafayette, IN 47907, USA}\columbia
\author{K.~E.~Lim}\email[Corresponding Author: ]{kelim@phys.columbia.edu}\columbia
\author{A.~J.~Melgarejo Fernandez}\columbia
\author{R.~Persiani}\bologna
\author{G.~Plante}\columbia
\author{A.~Rizzo}\columbia

\begin{abstract}
  We have measured the energy dependence of the liquid xenon (LXe)
  scintillation yield of electrons with energy between $2.1$ and
  $120.2 \1{keV}$, using the Compton coincidence technique. A LXe
  scintillation detector with a very high light detection efficiency
  was irradiated with ${}^{137}\n{Cs}$ $\gamma$ rays and the energy of
  the Compton-scattered $\gamma$ rays was measured with a high-purity
  germanium (HPGe) detector placed at different scattering angles. The
  excellent energy resolution of the HPGe detector allows the
  selection of events with Compton electrons of known energy in the
  LXe detector. We find that the scintillation yield initially
  increases as the electron energy decreases from $120 \1{keV}$ to
  about $60 \1{keV}$ but then decreases by about 30\% from $60
  \1{keV}$ to $2 \1{keV}$. The measured scintillation yield was also
  measured with conversion electrons from the $32.1 \1{keV}$ and $9.4
  \1{keV}$ transitions of the ${}^{83m}\n{Kr}$ isomer, used as an
  internal calibration source. We find that the scintillation yield of
  the $32.1 \1{keV}$ transition is compatible with that obtained from
  the Compton coincidence measurement. On the other hand, the yield
  for the $9.4 \1{keV}$ transition is much higher than that measured
  for a Compton electron of the same energy.  We interpret the
  enhancement in the scintillation yield as due to the enhanced
  recombination rate in the presence of Xe ions left from the $32.1
  \1{keV}$ transition, which precedes the $9.4 \1{keV}$ one by $220
  \1{ns}$, on average.
\end{abstract}



\maketitle

\section{Introduction} 

The experimental work presented in this paper is part of an ongoing
effort to understand the ionization and scintillation response of
liquid xenon (LXe) to low energy ($<$$10 \1{keV}$) particles, relevant
to the interpretation of data from dark matter searches based on LXe,
XENON100 in particular. Data from the current XENON100 experiment have
resulted in the most stringent limits to the interaction cross section
for a variety of dark matter Weakly Interacting Massive Particle
(WIMP) masses~\cite{Aprile:2010um,Aprile:2011hi,Aprile:2012nq}. The
next generation experiment, XENON1T, should provide almost two orders
of magnitude sensitivity improvement~\cite{Aprile:2012aa}.

The XENON detectors are time projection chambers in which both the
ionization, via proportional scintillation light, and the direct
scintillation light produced by particle interactions in the sensitive
LXe volume are recorded by photomultipliers
(PMTs)~\cite{Aprile:2011dd}. The scintillation and ionization response
of LXe depends on the electronic stopping power for the recoil type,
its energy, and the strength of the applied electric field. The
detector energy scale, for a given type of recoil, can in principle be
constructed from the scintillation signal, the ionization signal, or a
combination of both.  Inferring the energy of the particle from the
measured signals thus requires a precise knowledge of the response of
LXe to low-energy nuclear recoils, produced by WIMPs or background
neutrons, and electronic recoils, produced by electromagnetic
background. We have already reported several measurements of the
relative scintillation efficiency of nuclear recoils in
LXe~\cite{Aprile:2005mt,Aprile:2008rc,Plante:2011hw}, with the latest
measurements giving the most precise values to date for this quantity
and for recoil energies as low as 3 keV. The abundance of measurements
of the relative scintillation efficiency of nuclear recoils, compared
to the relatively few measurements of their ionization yield, is the
reason why a scintillation-based energy scale is often chosen instead
of an ionization-based or a ``combined'' energy scale. In this paper
we present our first measurement of the scintillation yield of
electronic recoils in the energy range of $2.1 \1{keV}$ to $120.2
\1{keV}$.

A recoiling electron in LXe produces a track of ionized and excited Xe
atoms or excitons. Both excitons and Xe ions that recombine with
electrons lead to the formation of excited dimers which subsequently
de-excite and produce scintillation photons~\cite{}. The ratio of the
number of excitons to the number of ions produced,
$N_{\n{ex}}/N_{\n{i}}$, is between 0.06 and 0.20~\cite{Doke:2002aa}
and hence the contribution to the scintillation signal from direct
excitation is small. If an electric field is applied, the fraction of
scintillation light that originates from recombining electron-ion
pairs is reduced. This fraction can thus be varied by changing the
applied electric field. However, even at zero electric field, not all
electrons recombine in a time scale practical for the collection of
the scintillation photons produced~\cite{Doke:1988aa}. In LXe, the
non-linearity in the scintillation signal from electronic recoils at
zero electric field is understood as being the result of the energy
dependence of the recombination probability.

Measurements of the scintillation yield of electrons of low energy
($\lesssim$$\,100\1{keV}$) in LXe are scarce. At these energies, in
most cases, scintillation light yield measurements have been carried
out with mono-energetic
sources~\cite{Barabanov:1987fj,Obodovskii:1994aa,Yamashita:2004} where
photoelectric absorption is the dominant interaction. One disadvantage
of using photo-absorbed $\gamma$ rays to measure the scintillation
yield is that multiple energetic electrons are produced as a result of
the photo-absorption: a photoelectron with an energy $E_\gamma-E_b$,
the incident $\gamma$ ray energy minus the electron binding energy,
and a host of de-excitation Auger electrons or X-rays photo-absorbed
afterwards. The scintillation yield obtained is then the convolution
of the distribution of electron energies produced with the
scintillation response of LXe to electrons instead of that of an
electron of that energy. On the other hand, a $\gamma$-ray Compton
scatter produces a single energetic electron with an energy very close
to $E_\gamma-E'_\gamma$, the incident $\gamma$ ray energy minus the
scattered $\gamma$ ray energy. This is because Compton scattering is
essentially equally probable for all atomic electrons instead of only
for those with significant binding energies, as is the case for
photoelectric absorption. Furthermore, the low-energy electromagnetic
background in a LXe dark matter detector is induced by
Compton-scattered high-energy $\gamma$-rays from the radioactivities
present largely in construction materials and the environment. A
second difficulty arising in measurements with external low-energy
$\gamma$ rays is the shallow penetration depth into the active volume
of the LXe detector.

Measurements of the scintillation yield of low-energy electrons in LXe
have also been performed via internal irradiation with conversion
electrons from the $^{83m}\n{Kr}$
isomer~\cite{Manalaysay:2009yq,Kastens:2009pa}. Despite solving the
problems of low-energy external sources, the extremely limited number
of isotopes that can be used for such irradiations prevents the
measurement of the scintillation yield over a continuous energy range.

The Compton coincidence technique, introduced by Valentine and
Rooney~\cite{Valentine:1994cs,Rooney:1996aa} and further improved by
Choong~\textit{et al.}~\cite{Choong:2008aa}, allows the measurement of
the electron response of scintillators at low energies.  This method
uses the energetic electrons produced by Compton-scattered $\gamma$
rays from a mono-energetic, high-energy source incident upon a
scintillation detector. If a $\gamma$ ray of energy $E_{\gamma}$
scatters in the scintillator, exits with energy $E'_\gamma$, and does
not interact anywhere else, the energy of the Compton electron
produced, $E_{\n{er}}$, is given by

\begin{align}
    E_{\n{er}} & = E_{\gamma} - E'_{\gamma} \\
    & =  E_{\gamma} - \frac{E_{\gamma}}{1+\frac{E_{\gamma}}{m_e c^2}(1 - \cos \theta)}
\end{align}
where $m_e$ is the electron mass, and $\theta$ is the scattering
angle. By using a second detector in coincidence with the
scintillation detector and measuring the energy of the scattered
$\gamma$ ray, it is possible to select nearly mono-energetic
electronic recoils from the continuous spectrum of Compton electrons
produced. By varying the angle at which the second detector is
positioned and the range of scattered $\gamma$ energies selected, one
can choose the energy at which the electron response is measured.

The experimental setup is described in Sec.~\ref{sec:setup}, the
calibration in Sec.~\ref{sec:calibration}, the Compton coincidence
measurements and data analysis in Sec.~\ref{sec:compton_coincidence},
and the response to mono-energetic $\gamma$ sources in
Sec.~\ref{sec:monoenergetic_sources}. The results are presented in
Sec.~\ref{sec:results}, followed by a discussion in
Sec.~\ref{sec:discussion}.

\section{Experimental Setup}
\label{sec:setup}

\begin{figure}[!htb]
\begin{center}
  \includegraphics[width=0.53\columnwidth]{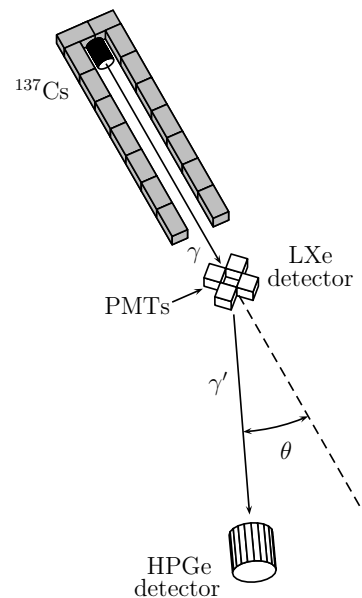}
  \caption{Schematic of the experimental setup. A $370 \1{MBq}$
    ${}^{137}\n{Cs}$ source is placed 85~cm from a LXe target viewed
    by six PMTs (only four shown, top and bottom PMTs are omitted for
    clarity). The energy of the $\gamma$ rays that scatter near an
    angle $\theta_\n{HPGe}$ are measured with a HPGe detector. The
    excellent energy resolution of the HPGe detector allows the
    selection of events where a Compton electron of the desired energy
    is produced in the LXe detector.}
\label{fig:setup}
\end{center}
\end{figure}

The measurement of the scintillation response of LXe to electronic
recoils was performed by irradiating a LXe detector with $\gamma$ rays
from a $370 \1{MBq}$ ${}^{137}\n{Cs}$ source and measuring the energy
of the scattered $\gamma$ rays at various angles with a high purity
germanium (HPGe) detector. Fig.~\ref{fig:setup} shows a schematic of
the experimental setup. The energy deposit in the LXe is inferred from
the energy measured in the HPGe detector. The scattering angle is
adjusted to select recoils in the desired energy range.

The LXe detector was designed with minimal materials outside of the
active volume to reduce the $\gamma$-ray scattering probability before
and after an interaction in the active volume. The active LXe volume
is a cube of 2.6~cm side covered by six 2.5~cm square Hamamatsu
R8520-406-SEL photomultiplier tubes (PMTs) mounted in a
polytetrafluoroethylene (PTFE) frame. The PMTs are the same type as
those used in the XENON100 experiment~\cite{Aprile:2011hi} but
selected for high quantum efficiency (QE). They have a bialkali
photocathode designed for low-temperature operation down to
$-110^\circ\n{C}$, and have an average room temperature QE of 32\% at
178~nm, the wavelength at which Xe scintillates~\cite{Jortner:1965aa}.
The measured QE values were provided by Hamamatsu. The high QE of the
PMTs and the large photocathode coverage of the arrangement yields a
very high light collection efficiency and thus enables a low energy
threshold.  The PMT were biased with positive high voltage to keep the
PMT metal body and photocathode at ground potential, thereby ensuring
that no electric field is present in the LXe active volume. More
details on the LXe detector can be found in
Ref.~\onlinecite{Plante:2011hw}.

The LXe detector vessel was filled with 1.82~kg of LXe, the amount
required for the liquid level to reach 1~cm above the active
volume. The total LXe mass in the active volume is 50~g. During
operation, the Xe is purified in the gas phase by circulating it
through a hot getter with a diaphragm pump.  The purified gas is
re-liquefied efficiently using a heat exchanger~\cite{Giboni:2011wx}.
The LXe temperature is kept constant with an Iwatani PDC08 pulse tube
refrigerator (PTR) delivering $24 \1{W}$ of cooling power at
$-106^\circ\n{C}$. More details on the cooling system for this
experiment are given in Ref.~\onlinecite{Giboni:2011wx}.  For the
measurements presented here, the LXe temperature was maintained at
$-95^\circ\n{C}$ which corresponds to a vapor pressure of
$2\1{atm}$. The LXe detector operating conditions were stable
throughout the entire data taking period with observed LXe temperature
and gaseous xenon (GXe) pressure variations (standard deviation over
mean) of less than 0.7\% and 0.6\%, respectively.

The Compton-scattered $\gamma$ rays were tagged with an ORTEC p-type
coaxial HPGe detector of 5.8~cm diameter and 4.8~cm depth. The typical
full width at half-maximum (FWHM) energy resolution at 1.33~MeV and
the peak-to-Compton ratio are specified by ORTEC to be less than
2.09~keV and better than 51:1, respectively.

The ${}^{137}\n{Cs}$ source was aligned with respect to the center of
the LXe detector active volume using an auto-leveling laser. The
desired HPGe detector floor positions were measured with a 1.5~m
aluminium rule and a plumb line. The vertical position of the HPGe
detector was set with the laser.  The location of ${}^{137}\n{Cs}$
source was fixed at distance of 85~cm from the center of the active
volume of the LXe detector. Lead bricks lined the path between the
source and the LXe detector to minimize the scattering of $\gamma$
rays outside the active volume of the LXe detector. The distance
between the LXe detector and the HPGe detector was varied from 14~cm
to 62~cm (see Table~\ref{tab:far_det_config}). The uncertainty in the
position of the HPGe detector was estimated to be less than 3~mm.

The signals from the six PMTs were fed into a Phillips 776 $\times 10$
amplifier with two amplified outputs per channel. The first output of
each channel was digitized by a 14-bit CAEN V1724 100~MS/s flash ADC
with 40~MHz bandwidth, while the second output was fed to a Phillips
706 leading edge discriminator. The discriminator thresholds were set
at a level of -20~mV, which corresponds to 0.7~photoelectrons
(pe). The logic signals of the six discriminator outputs were added
with a CAEN N401 linear fan-in and discriminated to obtain a twofold
PMT coincidence condition. The twofold PMT coincidence logic signal
was then passed to a $10\, \mu\n{s}$ holdoff circuit to prevent
re-triggering on the tail of the LXe scintillation signal, and
constituted the LXe trigger.

The signal of the HPGe detector was amplified with an ORTEC A257N
preamplifier and shaped with an ORTEC 450 research amplifier using
1~$\mu$s and 0.5~$\mu$s differentiation and integration time
constants, respectively.  The output of the research amplifier was
split with a passive resistive fan-out. One copy went directly to the
flash ADC and the other copy was discriminated at a threshold level of
-30~mV, to form the HPGe trigger signal.

Finally, for the Compton coincidence measurements presented here, the
trigger was given by the coincidence within a 200 ns window of the LXe
and the HPGe trigger signals.

The energy dependence of the efficiency of the LXe trigger was
measured using a ${}^{22}\n{Na}$ source and a NaI(Tl) detector with
the technique described in Ref.~\onlinecite{Plante:2011hw}. The result
obtained was compatible with the measurement of
Ref.~\onlinecite{Plante:2011hw}, confirming that recoil energy spectra
do not suffer efficiency losses in the energy region of interest. For
some of the data sets taken at higher energies ($\theta_\n{HPGe} =
8.6^{\circ}$, $16.1^{\circ}$), the threshold levels were set to -40~mV
so as to reduce the fraction of noise triggers. These increased
thresholds also did not decrease the event acceptance in the energy
region of interest.

\section{Calibration}
\label{sec:calibration}

\subsection{LXe Detector Calibration}
\label{subsec:lxe_chamber_calibration}

A blue light emitting diode (LED) embedded in the PTFE mounting
structure was used to calibrate the gain of each PMT. The light level
from the LED was adjusted such that the contamination of the
single-photoelectron peak from the double-photoelectron peak was
negligible. The gain value for each LED data set was determined by
fitting both the single-photoelectron peak and the noise pedestal with
Gaussian functions. The gain was taken as the difference between the
means of each Gaussian. The PMT gain calibration was performed at
regular intervals during data taking. For the analysis presented here,
the gain of each PMT was taken as their average measured gain over the
whole data taking period and its uncertainty as the variation in the
individual gain measurements. The uncertainty in the gain of each PMT
varied between 1\% and 1.6\%. Since the total scintillation signal is
obtained from the sum of all PMT signals, this leads to a total
contribution to the uncertainty on the measured scintillation signal
of 3\%.

\subsection{HPGe Detector Calibration}
\label{subsec:hpge_calibration}

The excellent energy resolution of the HPGe detector makes it possible
to select with high efficiency events where $\gamma$ rays Compton
scatter once and deposit a fixed energy in the LXe detector.  Since
the energy of the electronic recoil in the LXe detector is directly
determined by the measured energy in the HPGe detector, it is
important to verify the stability of the HPGe detector response
throughout the measurements.

The HPGe detector was calibrated through dedicated measurements with
the ${}^{137}\n{Cs}$ source between each Compton coincidence
measurement. The linearity of the energy calibration was verified with
511~keV $\gamma$ rays from a ${}^{22}\n{Na}$ source.

In addition, the stability of the HPGe detector calibration was
monitored during each Compton coincidence measurement via accidental
coincidence events. Accidental coincidence events from uncorrelated
LXe and HPGe triggers occur when two different $\gamma$ rays interact
in the LXe detector and the HPGe detector within the 200~ns
coincidence window time. Since the accidental coincidence HPGe energy
spectrum is essentially the same, albeit with a smaller rate, as an
energy spectrum taken with the HPGe trigger, the 661.7~keV full
absorption peak from ${}^{137}\n{Cs}$ $\gamma$ rays incident on the
HPGe detector can thus be used to monitor the stability of the
calibration (see Fig.~\ref{fig:compton_coincidence-9_deg}
(right)). The HPGe detector calibration factor was also corrected for
adjustments of the DC offset of the HPGe channel of the flash ADC. For
the Compton conicidence measurements presented here, the maximum
variation in the corrected HPGe calibration factor was 0.2\%. The
energy resolution at 661.7~keV, obtained via accidental coincidence
events, varied between 1.0 and 1.7~keV (see
Table~\ref{tab:far_det_config}). This variation is attributed to long
term changes ($<$$0.5\1{mV}$) in the baseline of the HPGe channel.
The effect of these small baseline changes could have been eliminated
by optimizing the amplifier gain to use the full dynamic range of the
FADC.

\section{Compton Coincidence Measurements}
\label{sec:compton_coincidence}

\subsection{Measured Electronic Recoil Distributions}
\label{subsec:measured_er_distributions}

Compton coincidence data sets were taken with the HPGe detector
positioned at eight different scattering angles, $\theta_\n{HPGe}$:
$0^\circ$, $5.6^\circ$, $8.6^\circ$, $12.0^\circ$, $16.1^\circ$,
$21.3^\circ$, $28.1^\circ$, and $34.4^\circ$, with LXe and HPGe
detector distances varying between 14~cm and 62~cm, resulting in
electronic recoil spectra with energies ranging from 2.0~keV to
122.2~keV. At each angle, a range of electronic recoil energies are
deposited in the LXe detector due to the angular acceptance of the LXe
target and that of the HPGe detector. Therefore, the HPGe detector
positions were chosen so as to obtain recoil energies covering the
above energy range with sufficient statistics.
Table~\ref{tab:far_det_config} lists the HPGe detector positions used
for each angle. In addition, a second $34.4^\circ$ data set was taken
with a different LXe and HPGe detector distance to investigate a
possible systematic effect on the measured scintillation yield from
the HPGe detector position. Finally, two data sets with different
trigger configurations were taken at $0^\circ$ to help study
background contributions at recoil energies below 5~keV, one with a
LXe detector trigger only, and one with a HPGe detector trigger only.

Since electronic recoils with a range of energies are accessible in
one measurement with the HPGe detector at a given position, and since
the energy resolution of the HPGe detector is much narrower than this
energy range, the scintillation response at many different recoil
energies can be extracted from a single data set.  Moreover, the
scintillation response at the same energy can be extracted from data
sets which have overlapping recoil energy ranges.

\begin{table}[htbp]
  \caption{HPGe detector positions, measured full absorption peak energy resolutions, and selected
    electronic recoil energy ranges for all Compton coincidence data sets. The variation
    of the measured resolution is discussed in Sec.~\ref{subsec:hpge_calibration}.}
	\label{tab:far_det_config}
	\begin{tabular*}{\columnwidth}{@{\extracolsep{\fill}} c c c c}
		\hline \hline
		$\theta_\n{HPGe}$ & HPGe Detector & HPGe Detector & $E_\n{er}$ Range \\
		 & Distance (cm) & Resolution (keV)  &(keV) \\
		\hline
		$0^\circ$ & $14$ & $1.4$ & $2.2 - 26.5$ \\
		$5.6^\circ$ & $60$ & $1.0$  & $2.0 - 12.9$ \\
		$8.6^\circ$ & $40$ & $1.0$  & $5.1 - 28.8$ \\
		$12.0^\circ$ & $40$ & $1.0$ &$10.0 - 27.2$ \\
		$16.1^\circ$ & $62$ & $1.3$ &$21.8 - 36.2$ \\
		$21.3^\circ$ & $40$ & $1.0$ &$33.9 - 60.2$ \\
		$28.1^\circ$ & $40$ & $1.1$ & $63.2 - 90.2$ \\
		$34.4^\circ$ & $19$ & $1.7$ &$77.2 - 122.2$ \\
		$34.4^\circ$ & $40$ & $1.0$ &$112.2 - 114.2$ \\
		\hline \hline
	\end{tabular*}
\end{table}

\begin{figure*}[htbp]
	\begin{center}
		\begin{tabular}{c c}
			\includegraphics[width=0.9\columnwidth]{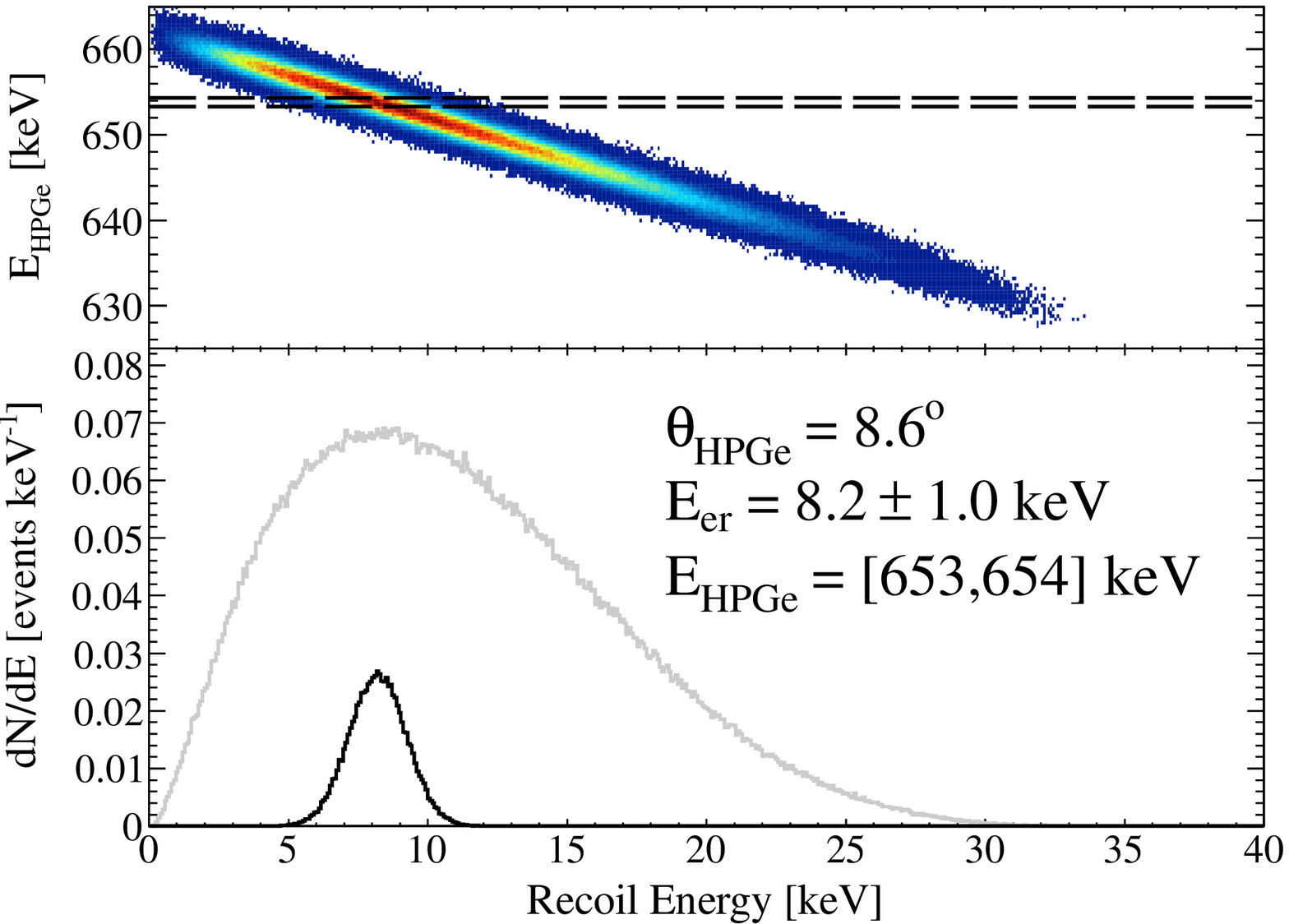} &
			\includegraphics[width=0.9\columnwidth]{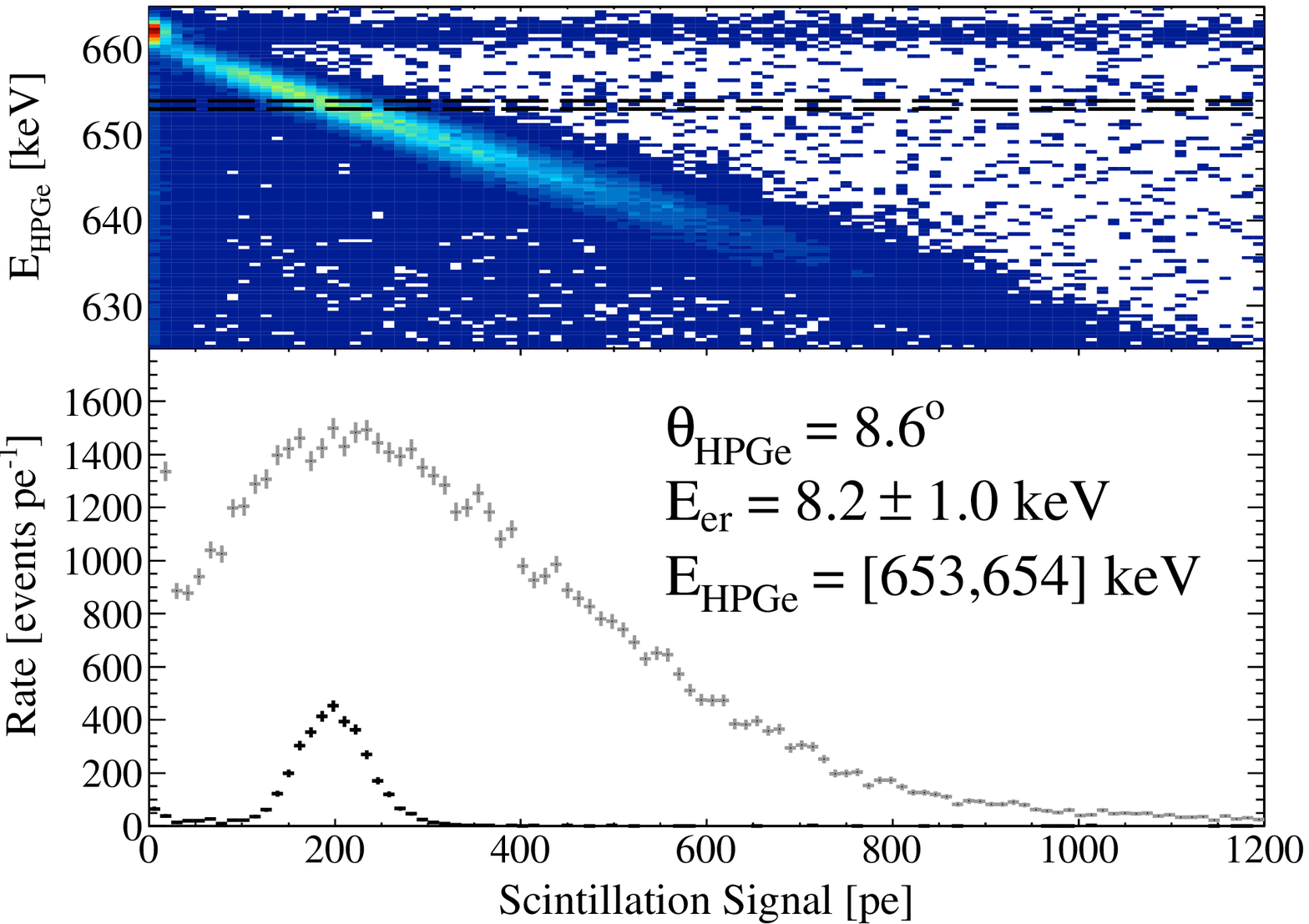} \\
		\end{tabular}
	\end{center}
	\caption{Simulated (left, top) and measured (right, top)
          distributions of HPGe detector energies and Compton electron
          recoil energies, or LXe scintillation signals in the case of
          the measurement, along with their projections (bottom, gray
          points) for the $8.6^\circ$ Compton coincidence setup. A
          known electronic recoil energy spectrum (black points) is
          obtained by selecting simulated events with HPGe detector
          energies between 653~keV and 654~keV (horizontal dashed
          lines). With this energy selection the spread in electronic
          recoil energies is dominated by the HPGe detector energy
          resolution of 1~keV at 661.7~keV measured for this dataset
          (Sec.~\ref{subsec:hpge_calibration}). Using the same energy
          selection (horizontal dashed lines), the scintillation
          response of LXe to 8.2~keV electronic recoils can be
          extracted from the $8.6^\circ$ Compton coincidence
          measurement. Additional backgrounds, neglected in the
          simulation, are present in the data. They become important
          only at recoil energies below 5~keV, as explained in the
          text.}
	\label{fig:compton_coincidence-9_deg}
\end{figure*}

The distribution of HPGe detector energies, $E_{\n{HPGe}}$, and
Compton electron recoil energies in the LXe detector, $E_{\n{er}}$,
for the $8.6^\circ$ Compton coincidence setup are shown in
Fig.~\ref{fig:compton_coincidence-9_deg}, for both data (right panel)
and a simplified Monte Carlo simulation (left panel). The distribution
of energy deposits in both detectors is shown in the top panel, while
the bottom one shows only the depositions in the LXe detector (gray
line). This simplified simulation, described in Sec.~\ref{subsec:mc},
only includes $\gamma$-ray interactions with the detector targets,
ignoring all other materials, and takes into account the energy
resolution of the HPGe detector. As expected, the energy of the
scattered $\gamma$ ray and that of the recoiling Compton electron sum
up to the energy of the $\gamma$ ray incident on the LXe detector,
$E_\gamma$.  Recoils over a range of energies are produced in the LXe
detector due to the angular acceptance of both detectors, as
expected. A distribution of known electronic recoil energies (black
line in the bottom panel) can be obtained by selecting a narrow range
of scattered $\gamma$-ray energies (horizontal dashed lines) measured
by the HPGe detector. The spread in electronic recoil energies after
the selection is given by the convolution of the energy range chosen,
$\Delta E_{\n{HPGe}}$, with the HPGe detector resolution near
$E_\gamma$. The scintillation response at a given electronic recoil
energy is obtained by calculating the mean scintillation signal
measured in the LXe detector when applying this HPGe detector energy
selection.

Fig.~\ref{fig:compton_coincidence-9_deg} (right, top), shows the
measured distribution of HPGe detector energies and LXe detector
scintillation signals for the $8.6^\circ$ Compton coincidence data
set. Comparing this with the distribution from the simulated data,
three different event populations are visible: events with $E_{\n{er}}
+ E_{\n{HPGe}}$ equal, lower, and higher than $E_{\gamma}$. The event
population where $E_{\n{er}} + E_{\n{HPGe}} = E_\gamma$, within the
limits of the HPGe detector resolution, corresponds to events where
the incident $\gamma$ ray scatters once in the active LXe volume,
producing a Compton electron of energy $E_{\n{er}}$, and is fully
absorbed in the HPGe detector. Consequently, the scintillation
response of LXe to nearly mono-energetic electronic recoils can be
inferred from these events.

The event population where $E_{\n{er}} + E_{\n{HPGe}}$ is lower than
$E_\gamma$ corresponds, for the most part, to events where the
scattered $\gamma$ ray deposits only a fraction of its energy in the
HPGe detector, due to the finite size of the crystal. That is, each
possible scattered $\gamma$-ray energy is responsible for a spectrum
of energies in the HPGe detector, with a full absorption peak, a
Compton continuum, a multiple Compton scattering region, the latter
two being responsible for the event population with $E_{\n{HPGe}}$
lower than the scattered $\gamma$-ray energy. Events where $\gamma$
rays scatter in other materials before interacting in the HPGe
detector additionally contribute to this population. A Monte Carlo
simulation based on the GEANT4 toolkit~\cite{Agostinelli:2002hh}, also
described in Sec.~\ref{subsec:mc}, was used to estimate the
contribution of such events in the energy range of the single scatter
peak for various electronic recoil spectra.

Finally, the event population where $E_{\n{er}} + E_{\n{HPGe}}$ is
higher than $E_\gamma$ corresponds to events with an accidental
coincidence between the LXe detector and the HPGe detector. This
population is especially pronounced at $E_{\n{HPGe}} \approx 661.7
\1{keV}$ in Fig.~\ref{fig:compton_coincidence-9_deg} (right), as
expected since the accidental coincidence spectrum should have a peak
at the incident $\gamma$-ray energy. As mentioned in
Sec.~\ref{subsec:hpge_calibration}, events from this population were
used to monitor the stability of the HPGe energy calibration during
the Compton coincidence measurements.

The increase in rate at low recoil energies compared to the simulated
data is attributed to events where the $\gamma$ ray interacts only in
the LXe outside the active volume but the resulting scintillation
light is visible in the active volume.  The feature is also observed
with all external $\gamma$-ray sources.  The average probability for a
photon outside the active LXe volume to reach a PMT photocathode was
estimated at $1 \times 10^{-4}$ via a light propagation Monte Carlo
simulation. An exponential feature consistent with that observed in
the data can also be reproduced in simulations by including the
expected scintillation signal from energy deposits outside the active
LXe volume. As is apparent from
Fig.~\ref{fig:compton_coincidence-9_deg} (right, top), the largest
background in the measurement of the scintillation response of LXe
with this technique is from accidental coincidences at low electronic
recoil energies.

\subsection{Monte Carlo Simulation}
\label{subsec:mc}

For optimum efficiency, two different Monte Carlo simulations were
used to analyze different aspects of the expected event distributions
for Compton coincidence measurements. The first is a simplified Monte
Carlo simulation that considers only events in which the incident
$\gamma$ ray interacts in the LXe detector, and deposits its full
energy in the HPGe detector. The second simulation is based on the
GEANT4 toolkit and includes a realistic description of the LXe
detector, detector vessel, vacuum cryostat, support frame, and HPGe
detector. It was used to obtain the expected electronic recoil energy
spectra as a function of HPGe energy, and thus enabled a direct
comparison with the measured spectra, and the identification and
quantification of the different backgrounds present.

The simplified Monte Carlo simulation incorporates the geometry of the
active LXe volume and of the HPGe detector crystal, the position of
the $^{137}\n{Cs}$ source, as well as the actual positions of the HPGe
detector used for the various Compton coincidence data sets. The
simulation proceeds by generating random positions within the volume
of the LXe detector, taking into account the Compton scattering mean
free path, and on the front surface of the HPGe detector, and then
calculating the recoil energy that corresponds to each pair of random
LXe and HPGe interaction points via the Compton scattering
formula. The energy deposited in the HPGe detector is then simply
taken as the incident $\gamma$-ray energy, $E_\gamma$, minus the
recoil energy in the LXe detector, thus assuming that the scattered
$\gamma$ ray deposited its full energy in the HPGe detector. This is
then convolved with a Gaussian energy resolution.  The standard
deviation used for each Compton coincidence data set is the value
measured using the corresponding accidental coincidence spectrum (see
Sec.~\ref{subsec:hpge_calibration}). Calculating the expected recoil
energy from this simulation assumes that the incident $\gamma$ ray
travels directly from the source to the LXe detector, scatters once in
the LXe detector, and travels directly to the HPGe detector, thereby
neglecting any interactions in materials outside of the LXe active
volume. Furthermore, since scattering angles are not sampled from the
photon differential scattering cross-section, the calculation neglects
any angular dependence in the cross section over the range of
scattering angles geometrically allowed by both
detectors. Nevertheless, the expected mean energy of the recoil peak
from the simplified simulation was found to be in agreement at the 1\%
level with that of the GEANT4-based simulation.  In addition, the
simulated spectra agree with each other at all recoil energies above 2
keV. Disagreement on the order of 10\% appears for the 2 keV recoil
peak below 1 keV.

As mentioned earlier, the resulting mean and spread of the electronic
recoil peak in the LXe detector, for each HPGe energy selection window
applied to a Compton coincidence data set, were calculated using the
simplified simulation by applying the appropriate energy selections to
each simulated data set. The effect of the misalignment of the HPGe
detector on the mean energy of the recoil peak was investigated by
varying the position of the HPGe detector in the simulation. Mean
recoil energies are found to vary by less than 2\%. Finally, the
change in the response of the LXe detector due to the variation of the
spatial event distribution in the LXe with the HPGe energy selection
was estimated by calculating the average light detection efficiency
over the spatial distribution of events for different HPGe energy
selections. The spatial variation of the light detection efficiency
used for the calculation was obtained from an independent light
propagation Monte Carlo simulation. This simulation takes into account
the geometry of the PMTs and the PTFE holding structure,the
reflectivity of the materials in contact with the active LXe volume,
the QE and collection efficiency of the PMTs, and an estimate of the
angular response of the PMTs~\cite{hamamatsu:pmt_handbook}.

The GEANT4-based Monte Carlo simulation uses the same description of
the LXe detector as the one used to simulate the expected nuclear
recoil energy distributions for the measurement of the scintillation
efficiency of low-energy nuclear recoils in LXe that was performed
with the same detector~\cite{Plante:2011hw}.  The geometry and
response of the HPGe detector was verified by comparing simulated
energy spectra with measured spectra from dedicated ${}^{137}\n{Cs}$
calibrations of the HPGe detector. The information recorded in the
simulation includes the energy, position, time, type of particle and
physical process responsible for each energy deposit in the LXe
detector, as well as the total energy, time, and type of particle for
each energy deposit in the HPGe detector.

\begin{figure*}[htbp]
	\begin{center}
		\begin{tabular}{c c}
			\includegraphics[width=0.9\columnwidth]{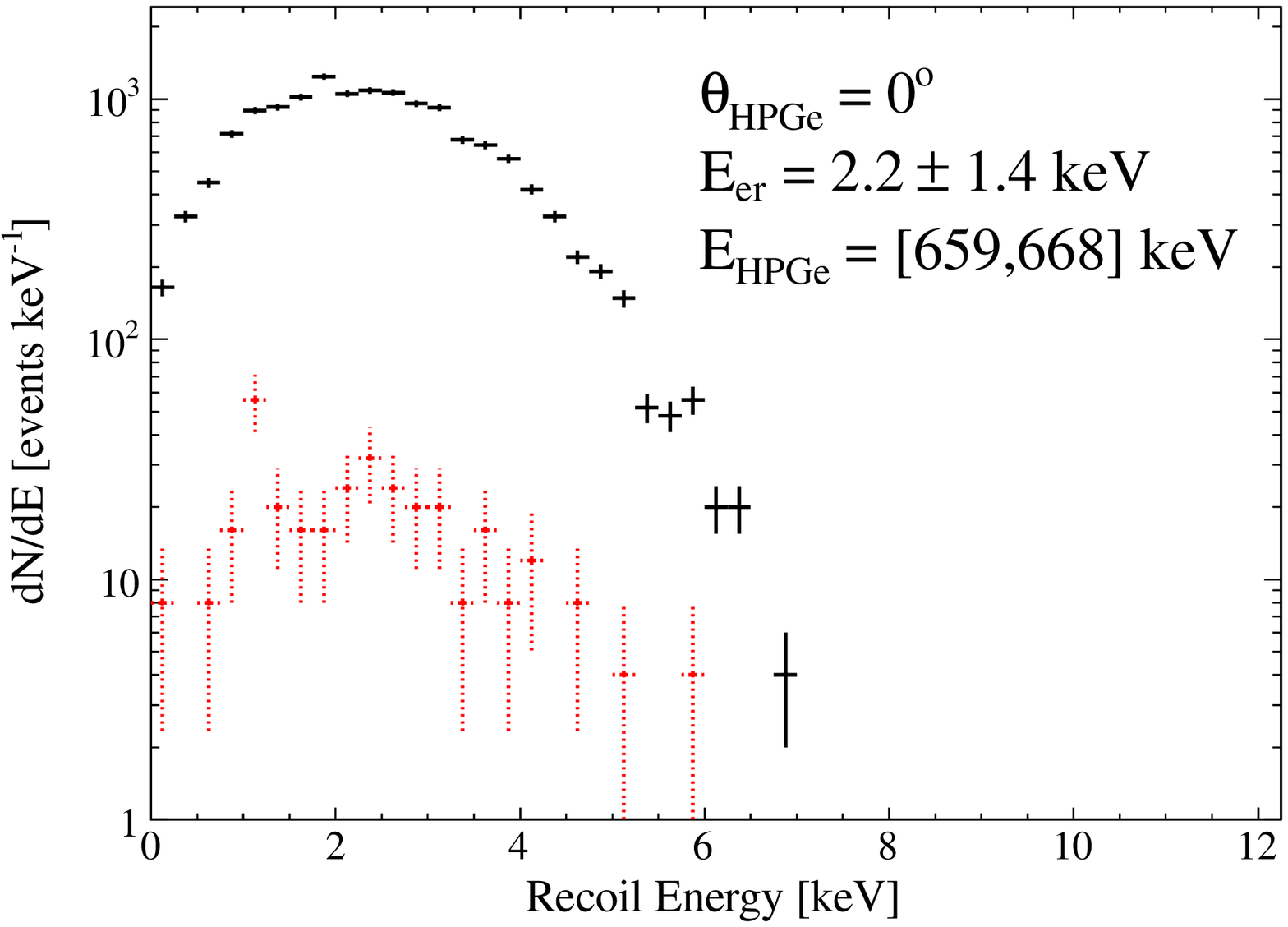} &
			\includegraphics[width=0.9\columnwidth]{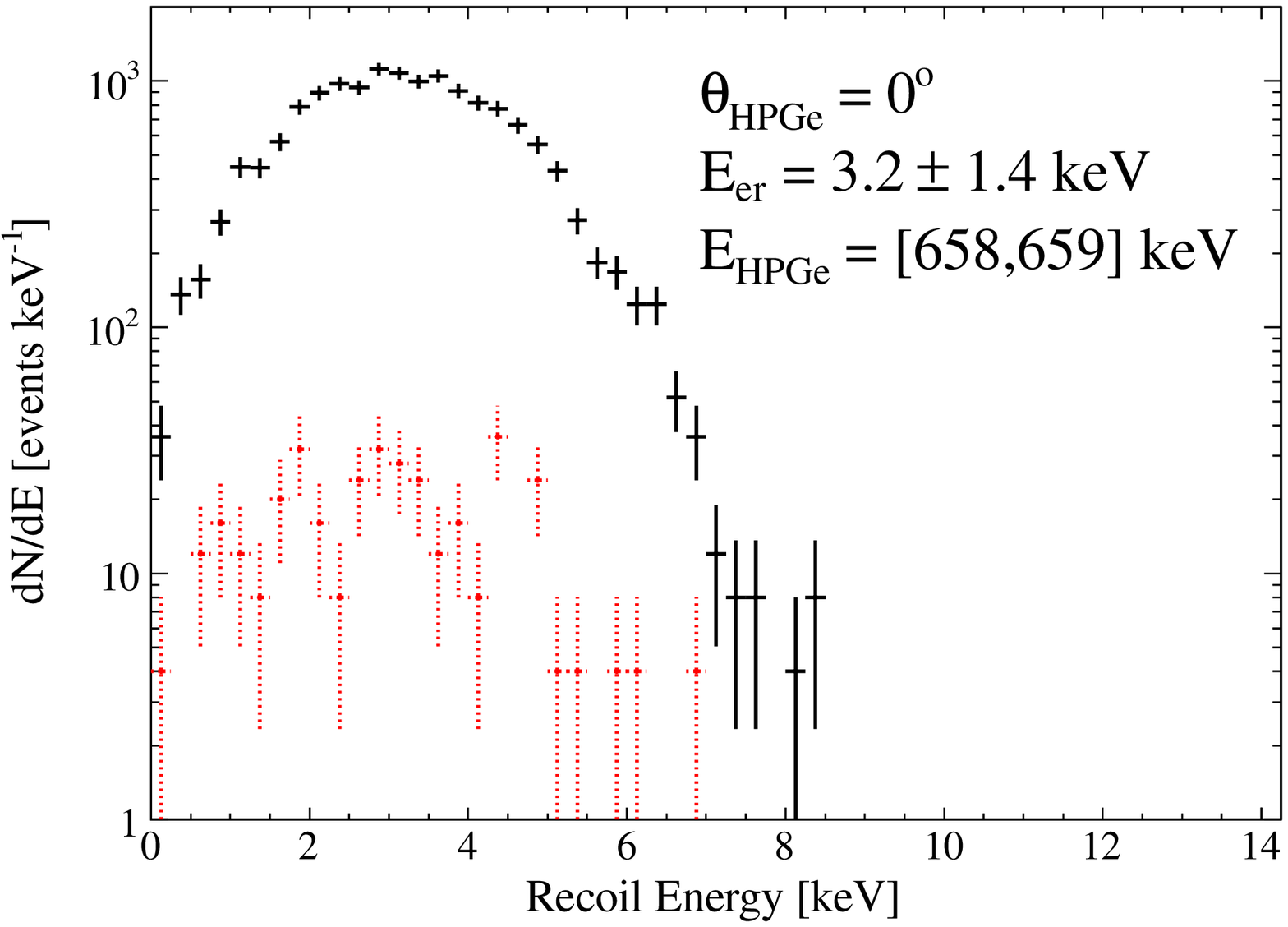} \\
			\includegraphics[width=0.9\columnwidth]{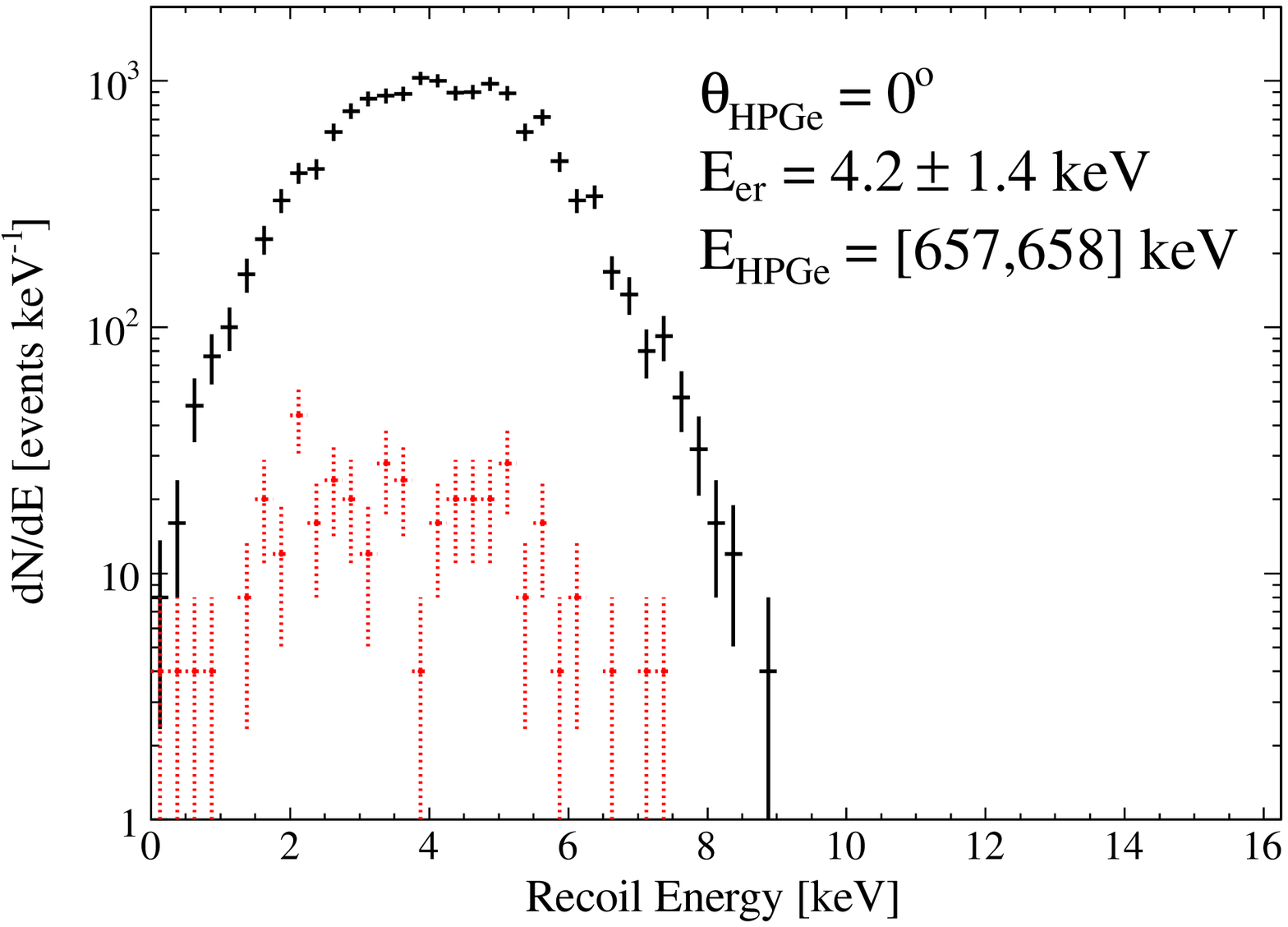} &
			\includegraphics[width=0.9\columnwidth]{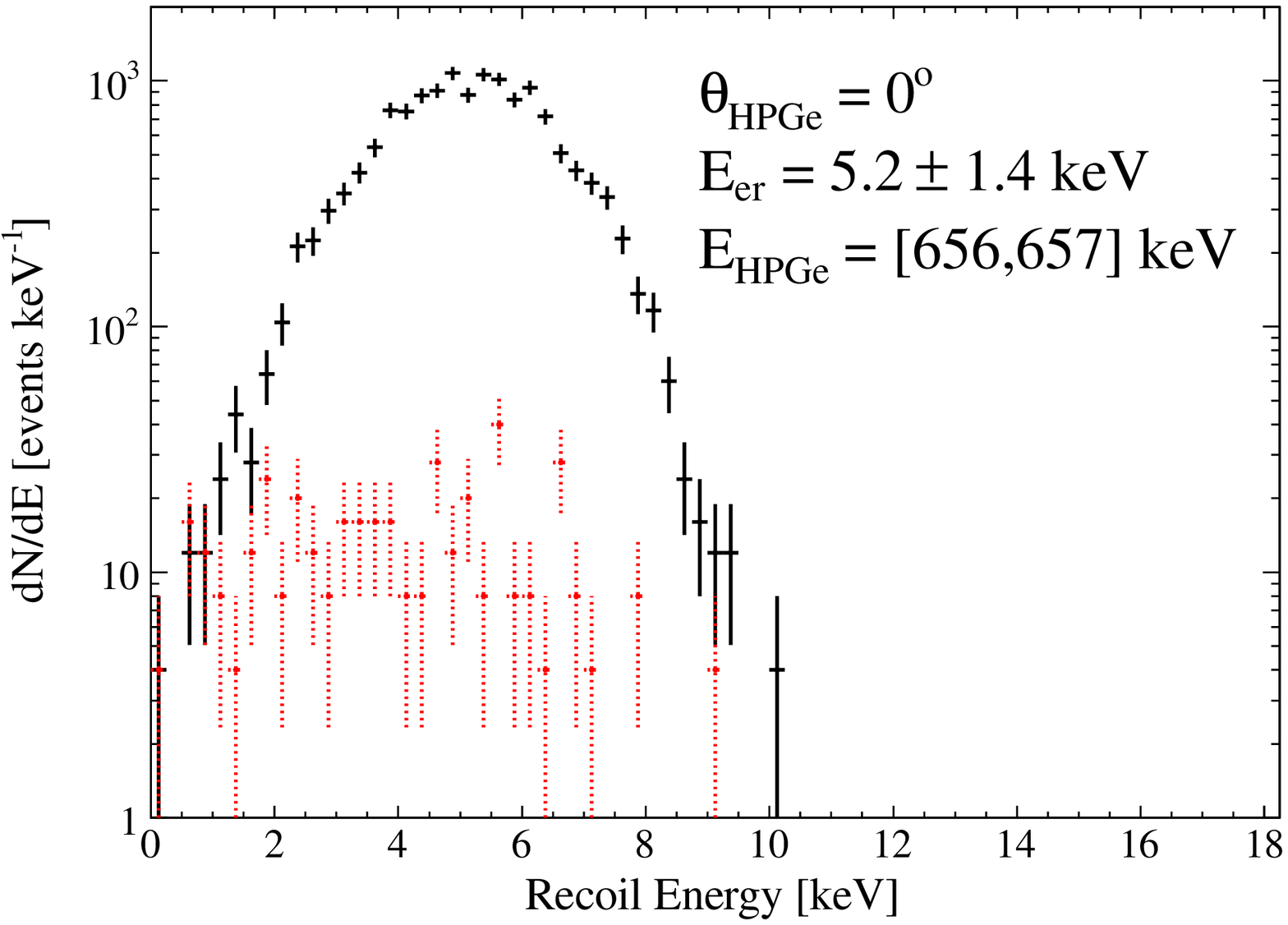} \\
		\end{tabular}
	\end{center}
	\caption{Simulated electronic recoil energy spectra for the
          $0^\circ$ Compton coincidence setup, using $E_\n{HPGe}$
          energy selections $\left[{659,660}\right]$,
          $\left[{658,659}\right]$, $\left[{657,658}\right]$, and
          $\left[{656,657}\right]$~keV, resulting in mean recoil
          energies of $2.2 \pm 1.4$, $3.2 \pm 1.4$, $4.2 \pm 1.4$ and
          $5.3 \pm 1.4 \1{keV}$, respectively.  The black histograms
          show the spectrum of events where the incident $\gamma$ ray
          interacts in the active LXe volume, and nowhere else, and
          deposits in the HPGe detector an energy within the HPGe
          selection window. The red histogram corresponds to events
          where the $\gamma$ ray additionally interacts in other
          materials, either before or after scattering in the active
          LXe volume, before interacting in the HPGe detector. The
          contamination of the recoil peak by events with $\gamma$-ray
          interactions in other materials is less than
          3\%.}
	\label{fig:mc-0deg_low_kev}
\end{figure*}

Fig.~\ref{fig:mc-0deg_low_kev} shows the simulated electronic recoil
energy spectra for the $0^\circ$ Compton coincidence setup, obtained
from the GEANT4-based simulation using $E_\n{HPGe}$ energy selections
$\left[{659,660}\right]$, $\left[{658,659}\right]$,
$\left[{657,658}\right]$, and $\left[{656,657}\right]$~keV, resulting
in mean recoil energies of $2.2 \pm 1.4$, $3.2 \pm 1.4$, $4.2 \pm 1.4$
and $5.3 \pm 1.4 \1{keV}$, respectively. The black spectra consist of
events in which the $\gamma$ scattered nowhere else than in the active
LXe volume before interacting in the HPGe detector whereas the red
spectra consist of events in which the $\gamma$ ray additionally
interacted in other materials, either before or after scattering in
the active LXe volume, before interacting in the HPGe detector. The
contribution of these multiple scatter events to the electronic recoil
peak is less than 3\%.  Their energy spectrum is not peaked since the
presence of additional scatters spoils the HPGe energy and LXe recoil
energy correlation. Note, however, that since the selection is for a
fixed HPGe energy, the maximum recoil energy for these events is
constrained to be lower than the maximum energy of the recoil
peak. Multiple scatters in the active LXe volume are highly suppressed
due to the small size of the target with respect to the Compton
scattering mean free path in LXe for ${}^{137}\n{Cs}$ $\gamma$ rays
($\sim$$5.5 \1{cm}$).  These spectra can be compared to the measured
LXe scintillation spectra shown in Fig.~\ref{fig:0deg_low_s1tot},
keeping in mind that the background contribution from accidental
coincidence events is not included in the simulation. At energies of
3.2~keV and above, the measured electronic recoil peak is well
separated from the background from accidental coincidences. This low
contamination from events with scatters in other materials shows that
the design goal of minimizing the amount of materials in the vicinity
of the active LXe volume has been achieved, in agreement with
Ref.~\onlinecite{Plante:2011hw}.

The electronic recoil spectra with mean recoil energies of $2.2 \pm
1.4$, $3.2 \pm 1.4$, $4.2 \pm 1.4 \1{keV}$ were also used to calculate
the uncertainty in the LXe scintillation response at low energies
arising from the assumption of an exponential background model
(Sec.~\ref{subsec:extracting_ly}). The details of the calculation are
described in Sec.~\ref{sec:results}.

\subsection{The Scintillation Yield}
\label{subsec:extracting_ly}

For each scattering angle ($\theta_\n{HPGe}$) at which Compton
coincidence measurements were taken, the distribution of HPGe detector
energies and LXe scintillation signals were divided in $1 \1 {keV}$
slices along the $E_{\n{HPGe}}$ axis and the resulting LXe
scintillation spectra were analyzed for each of the selected energies.

\begin{figure*}[htbp]
	\begin{center}
		\begin{tabular}{c c}
			\includegraphics[width=0.9\columnwidth]{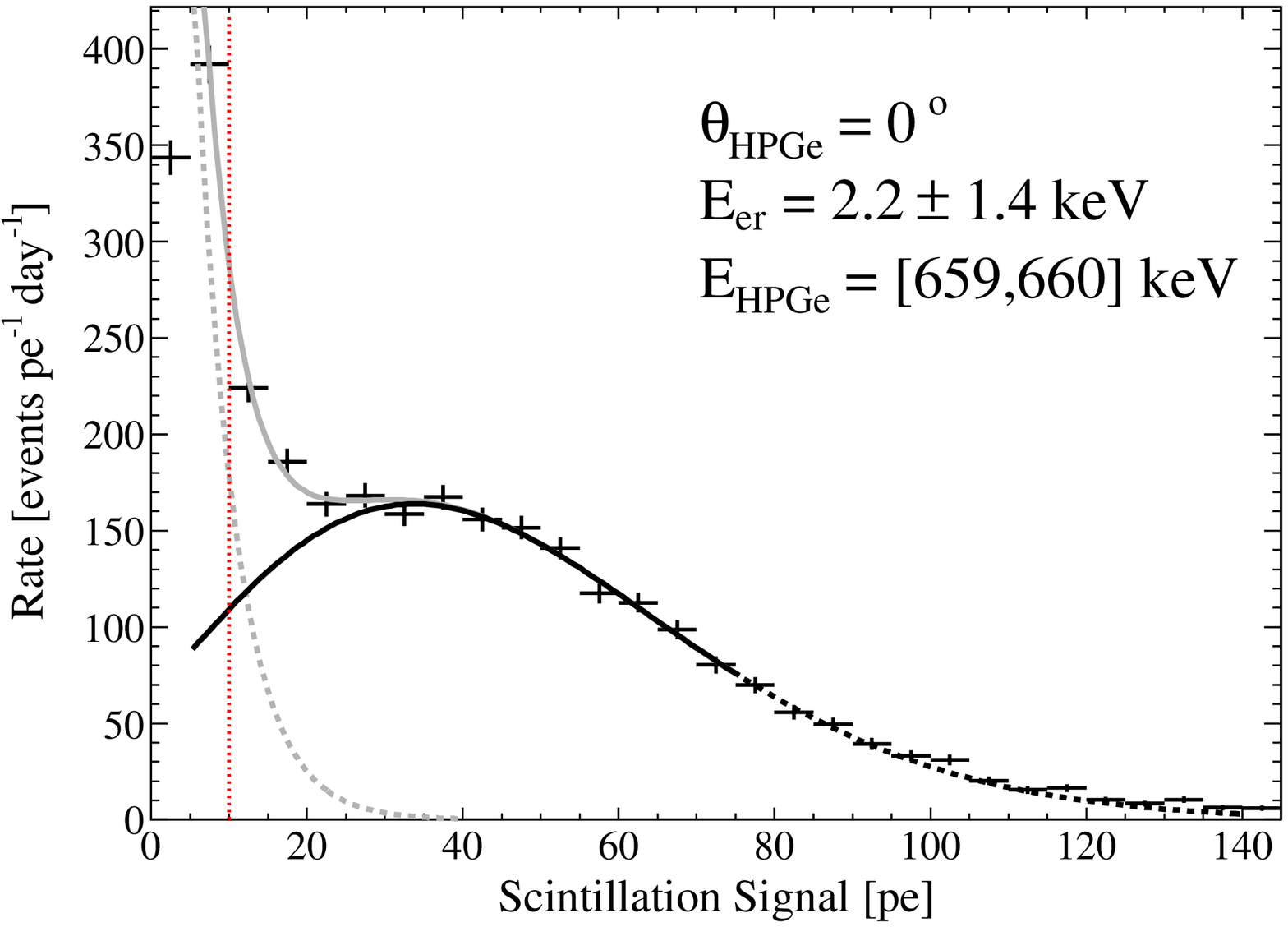} &
			\includegraphics[width=0.9\columnwidth]{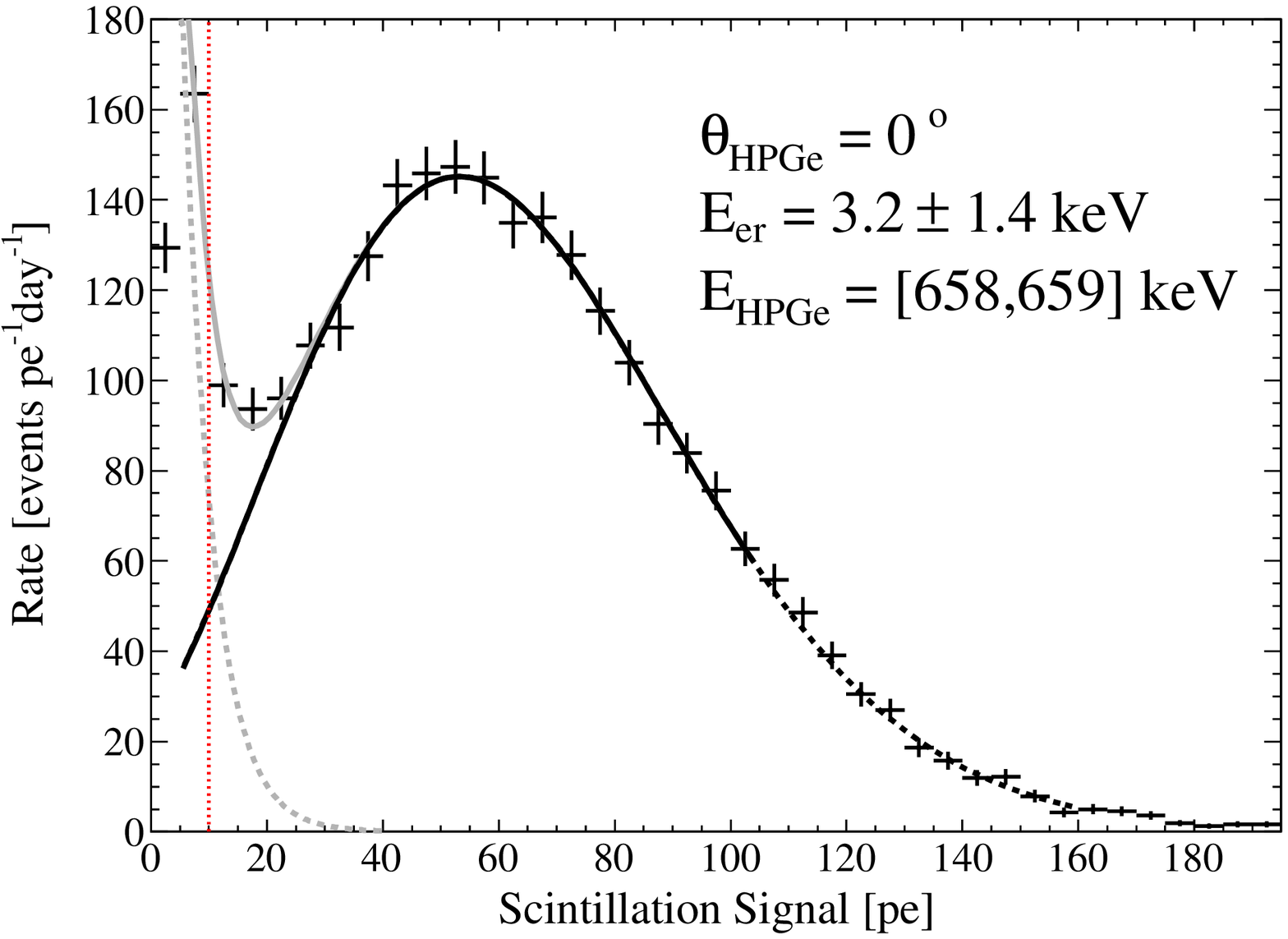} \\
			\includegraphics[width=0.9\columnwidth]{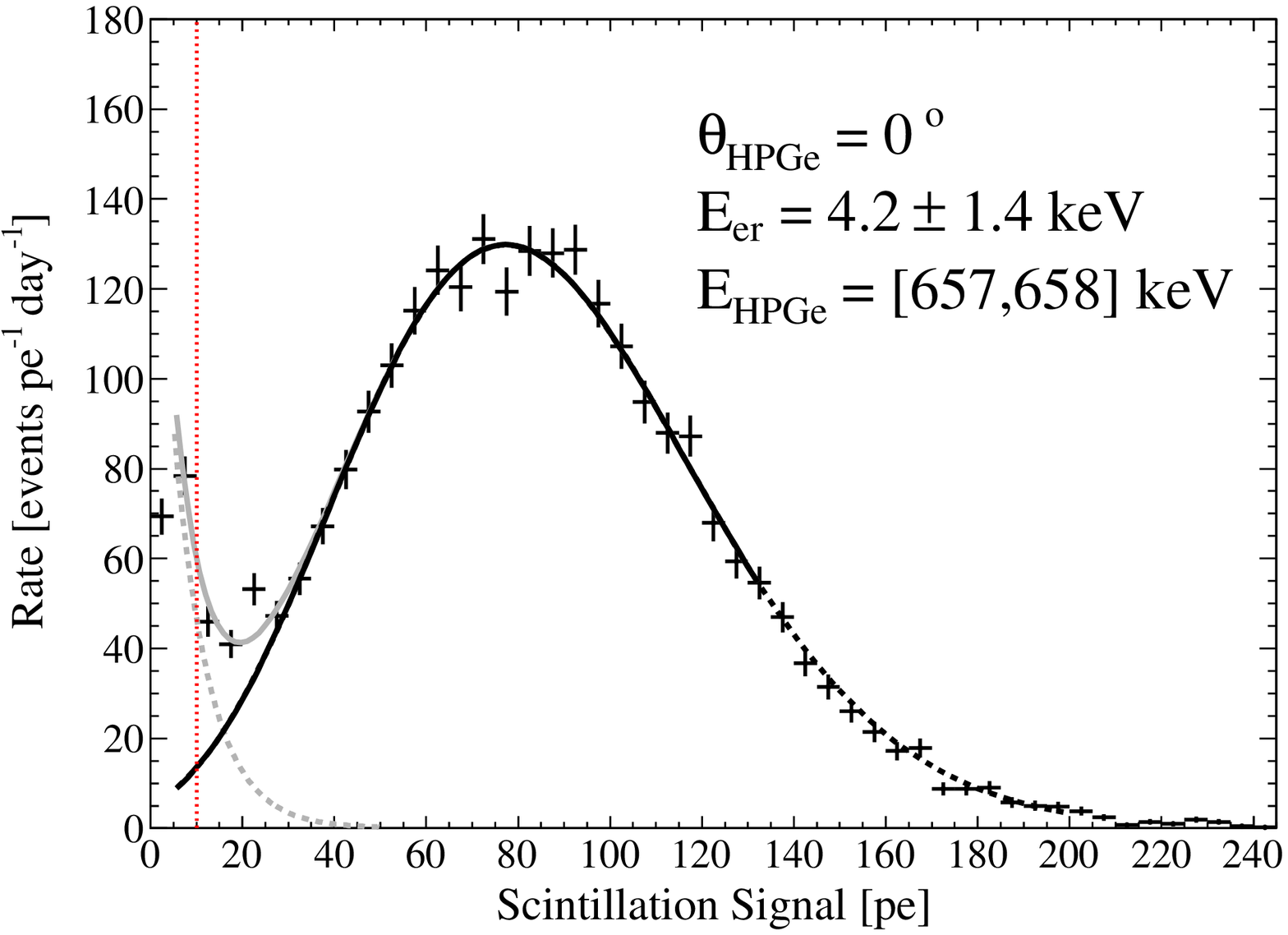} &
			\includegraphics[width=0.9\columnwidth]{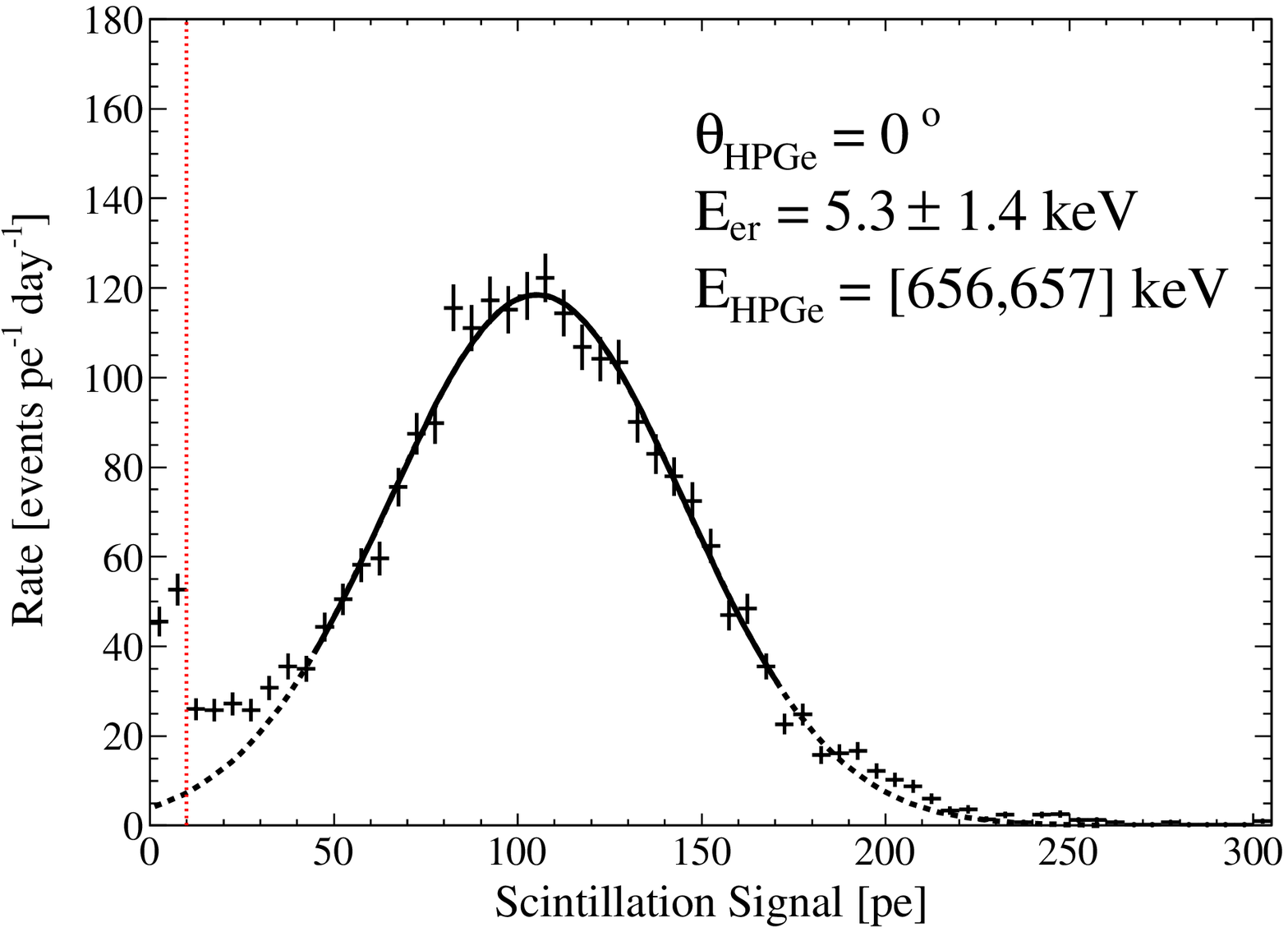} \\
		\end{tabular}
	\end{center}
	\caption{LXe scintillation spectra (points) for electronic
          recoil energies of $2.2 \pm 1.4$, $3.2 \pm 1.4$, $4.2 \pm
          1.4$ and $5.3 \pm 1.4 \1{keV}$ from the $0^\circ$ Compton
          coincidence data set with the same HPGe energy selection
          windows used in the Monte Carlo analysis.  As a reference,
          the measured $>$$99\%$ trigger efficiency is indicated by
          the vertical red dashed line. For recoil energies $E_\n{er}$
          below 5~keV, the scintillation spectra were fitted with the
          sum (gray line) of a ``scaled'' continuous Poisson function
          (black points) and an exponential function (dashed
          gray line), as described in the text. The range used for
          each fit is indicated by the extent of the solid black
          line. For recoil energies above 5~keV, the spectra were
          fitted with Gaussian functions (black line).}
	\label{fig:0deg_low_s1tot}
\end{figure*}

Fig.~\ref{fig:0deg_low_s1tot} shows the LXe scintillation spectra
obtained for the four lowest electronic recoil energies from the
$0^\circ$ Compton coincidence data set. For recoil energies below 2
keV, the background in the signal region is too high to extract the
scintillation yield in LXe. The spectra consist of a recoil peak,
which corresponds to events where the incident $\gamma$ ray scattered
in the active LXe volume only and deposited its full energy in the
HPGe detector, and different backgrounds depending on the electronic
recoil energy range selected. For spectra at low recoil energies, the
background mostly comes from accidental coincident events from the
full absorption peak of $^{137}\n{Cs}$ in the HPGe detector and few
photoelectrons scintillation signals from the LXe detector, believed
to originate from interactions in the LXe outside the active volume,
as discussed earlier. For spectra at recoil energies above 5~keV, the
background largely comes from events in which scattered $\gamma$ rays
with energies higher than that expected for the HPGe energy selection
deposit only a fraction of their energy in the HPGe detector,
resulting in an approximately flat background from zero to the recoil
peak. Ultimately, spectra at recoil energies above or below the range
of energies expected from the angular acceptance of the LXe and HPGe
detectors are dominated by events where the $\gamma$ ray scattered in
other materials and by accidental coincidence events between a partial
energy deposit in both detectors.

For spectra at recoil energies below 5~keV, the recoil peaks are
slightly asymmetric, exhibiting a longer tail at higher
energies. Additionally, the background from accidental coincidence
events is significant and must be taken into account to obtain the
correct LXe scintillation response. Consequently, the spectra were
fitted with the sum of a ``scaled'' continuous Poisson function, that
is, a function of the form $f_{\mu, a}\!\left({x}\right) = e^{-\mu}
\mu^{ax}/\Gamma\left({ax+1}\right)$, which describes the asymmetry of
the recoil peak with the scaling parameter $a$, and an exponential
function, which represents the background coming from accidental
coincidence events. Fig.~\ref{fig:0deg_low_s1tot} (top left, top
right, bottom left) shows the results of fits to spectra at electronic
recoil energies of $2.2 \pm 1.4$, $3.2 \pm 1.4$, and $4.2 \pm 1.4
\1{keV}$ from the $0^\circ$ Compton coincidence data set,
respectively. Note that the uncertainty on the electronic recoil
energy stated here (and throughout) corresponds to the spread in
recoil energies after the HPGe energy selection (see
Fig.~\ref{fig:0deg_low_s1tot} left), which is dominated by the HPGe
energy resolution, and not the uncertainty on the mean energy of the
recoil peak, which is considerably smaller.

For spectra at recoil energies above 5~keV, the recoil peaks are
symmetric and the fraction of events arising from background is
small. Hence these spectra were fitted with Gaussian functions over
the range of the recoil peaks. Fig.~\ref{fig:0deg_low_s1tot} (bottom
right) shows the result at $5.3 \pm 1.4 \1{keV}$ from the $0^\circ$
Compton coincidence data set. The background from scattered $\gamma$
rays with partial energy deposited in the HPGe detector is apparent to
the left of the recoil peak.

As explained in Sec.~\ref{subsec:measured_er_distributions}, each
Compton coincidence data set can be used to infer the scintillation
response over a wide range of energies, limited mostly by the angular
acceptance of the LXe and HPGe detectors at the position used for each
measurement. For recoil energies near the extremes of the range of
energies for a given configuration, the background from multiple
scatter and accidental coincident events dominates over the recoil
peak. The range of electronic recoil energies over which the
scintillation response was calculated was chosen for each data set so
that the fraction of events attributable to background in the recoil
peak would remain below 20\%. To estimate the background contribution
in the recoil peak, the event rate in the regions between 2 and
4~$\sigma$ above and below the peak was computed. This value was then
scaled to the width of the peak fitting range and divided by the total
event rate in this range. This background contamination estimation
method is valid as long as the background varies smoothly in energy,
as was observed to be the case in all spectra above recoil energies of
5~keV.  Table~\ref{tab:far_det_config} lists the resulting ranges over
which the scintillation response was calculated for each Compton
coincidence data set.

The mean electronic recoil energy does not exactly correspond to
$E_\gamma$ minus the central value of the HPGe energy range selected,
because the event rate varies as a function of the recoil energy (see
Fig.~\ref{fig:compton_coincidence-9_deg}), due to the geometrical
acceptance of the detectors. In a region where the event rate
increases as a function of recoil energy, for $\gamma$-ray scattering
angles smaller than the angle at which the HPGe detector is
positioned, the mean electronic recoil energy obtained from the HPGe
energy selection will be higher than expected since more events at
higher recoil energies will be included in the selection. The finite
energy resolution of the HPGe detector accentuates this effect since
even more events from higher or lower energies will be shuffled. The
mean and spread of the electronic recoil energy for a given HPGe
energy selection was calculated using the simplified Monte Carlo
simulation described in Sec.~\ref{subsec:mc}, applying the same energy
selection criteria as for the data.

The HPGe energy selection also has an effect on the spatial
distribution of events within the LXe detector.  Events for which the
$\gamma$-ray scattering angle is close to $\theta_\n{HPGe}$, and hence
those for which the HPGe energy selection window is close to
$E_\n{er}(\theta_\n{HPGe})$, will be distributed somewhat uniformly in
the center of the LXe detector. As the central value of HPGe energy
selection is decreased, however, the events will progressively cluster
near the side of the LXe detector towards higher scattering
angles. Similarly, events will progressively cluster near the side of
the LXe detector towards lower scattering angles when the central
value of the HPGe energy selection is increased.  The relative bias in
the measured scintillation response from this effect was estimated
using the simplified Monte Carlo simulation and found to be smaller
than 0.7\%, mostly due to the small spatial variation of the light
detection efficiency of the LXe detector~\cite{Plante:2011hw}. This
effect is further suppressed since the energy range over which the
scintillation response is calculated is already restricted by limiting
the maximum background contamination of the electronic recoil energy
peak. Recoil energy ranges corresponding to highly clustered event
distributions are thus avoided.

\section{Scintillation Response to Mono-Energetic Sources}
\label{sec:monoenergetic_sources}

Several radioactive sources were used to evaluate the response of the
LXe detector. Specifically, ${}^{137}\n{Cs}$, ${}^{22}\n{Na}$, and
${}^{57}\n{Co}$ external sources were used to obtain the $\gamma$-ray
response of the LXe detector while ${}^{83m}\n{Kr}$ was used as an
internal source for the response to fast electrons.

\subsection{Response to External $\gamma$-ray Sources}
\label{subsec:gamma_ray_sources}

The measurements with external sources were performed by attaching the
sources to the cryostat vessel at the height of the LXe active
volume. These measurements where taken without the additional $\times
10$ amplification (Sec.~\ref{sec:setup}) of the PMT signal to prevent
saturation of the flash ADC, which has a maximum input voltage of
2.25~V. In the normal configuration, saturation starts to occur for
signals of $10^3 \1{pe}$ on a single PMT whereas in this configuration
the response from the 1.275~MeV $\gamma$ ray from ${}^{22}\n{Na}$,
with a mean signal per PMT of $4.6 \times 10^3 \1{pe}$, could be
measured without any saturation effect.

\begin{figure}[!htb]
\begin{center}
	\includegraphics[width=1.0\columnwidth]{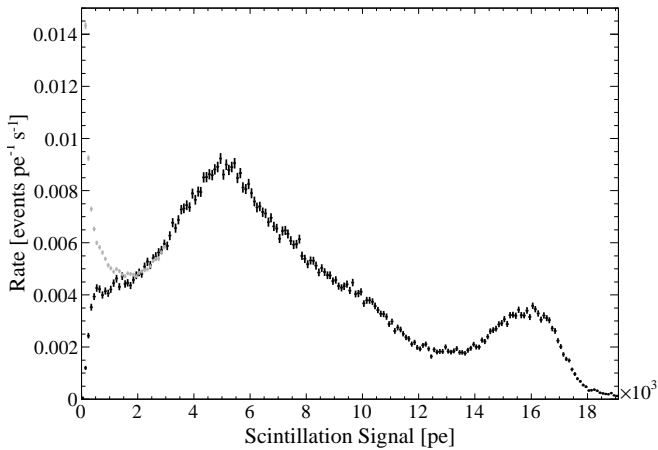}
	\caption{LXe scintillation spectrum obtained with the $370
          \1{MBq}$ ${}^{137}\n{Cs}$ $\gamma$ source without (black)
          and with (gray) additional $\times 10$ amplification.  The
          peak at $16 \times 10^3 \1{pe}$ corresponds to the 661.7~keV
          full absorption peak while the other peaked feature at $5
          \times 10^3 \1{pe}$ is mainly due to the backscatter peak.
          The event rate increase at low energies is visible in the
          spectrum with additional amplification, as is also observed
          in the accidental coincidence spectra from the Compton
          coincidence measurements.  }
\label{fig:cs137_s1}
\end{center}
\end{figure}

Fig.~\ref{fig:cs137_s1} shows a scintillation spectrum obtained with
the ${}^{137}\n{Cs}$ source. The peak at $16 \times 10^3 \1{pe}$
corresponds to the 661.7~keV full absorption peak while the other peaked
feature at $5 \times 10^3 \1{pe}$ is the backscatter peak, mainly due
to $\gamma$ rays that scatter in materials immediately surrounding the
LXe active volume before photoelectric absorption in the outer layers
of the active volume. The roll-off at low energies is due to the
increased effective trigger threshold when the additional $\times 10$
amplification is not applied to the PMT signals (black points). At low
energy, in the spectrum with the additional $\times 10$ amplification,
the event rate rises exponentially (gray points). As discussed in
Sec.~\ref{subsec:measured_er_distributions} the suspected origin of
these events is the small probability for scintillation photons
produced outside the active LXe volume to leak into it. This feature
at low energy is observed in all spectra obtained with external
$\gamma$-ray sources.

The large photocathode coverage and the use of PTFE as scintillation
light reflector on the few remaining surfaces assures a good
uniformity of the light collection efficiency throughout the active
volume. Even so, there is a slight increase in the light collection
efficiency near the surface of the PMT windows.  The light propagation
simulation mentioned in Sec.~\ref{subsec:mc} estimates this increase
to be $\sim$$6\%$ with respect to the volume-averaged light collection
efficiency. This LXe detector spatial non-uniformity can
systematically increase the measured scintillation yield of low-energy
$\gamma$ rays from external sources such as ${}^{57}\n{Co}$. To
mitigate this effect, three cuts on the relative light ratio between
two opposing PMTs are applied to the ${}^{57}\n{Co}$ data to select
interactions that occur further from the PMT windows. The
volume-averaged scintillation yield obtained at 122~keV is $23.60 \pm
0.03 \, \n{(stat)} \pm 0.85 \, \n{(sys)} \1{pe/keV}$, consistent with
the value of Ref.~\onlinecite{Plante:2011hw}.

Table~\ref{tab:source_ly} lists the measured scintillation yields for
the various external $\gamma$-ray sources used to evaluate the
scintillation response of the LXe detector.  The statistical
uncertainty comes from the fit of the spectra and the variation with
different fitting ranges on the spectra. The systematic uncertainty
includes contributions from the measured variations in the PMT gains
and in the response at different source positions.

\begin{table}[htbp]
	\caption{Measured scintillation yields for various external $\gamma$-ray sources and for the internal
		irradiation with ${}^{83m}\n{Kr}$.} 
	\label{tab:source_ly}
	\begin{tabular*}{\columnwidth}{@{\extracolsep{\fill}} l c c c}
          \hline \hline
          Source & Energy (keV) & Type & Scintillation Yield (pe/keV)\\
          \hline
          ${}^{22}\n{Na}$  & 1274.6 & $\gamma$ & $22.26\pm 0.08 \, \n{(stat)} \pm 0.77 \, \n{(sys)}$  \\
          ${}^{137}\n{Cs}$ & 661.7  & $\gamma$ & $23.84\pm 0.08 \, \n{(stat)} \pm 0.85 \, \n{(sys)}$ \\
          ${}^{22}\n{Na}$  & 511    & $\gamma$ & $23.76\pm 0.18 \, \n{(stat)} \pm 1.07 \, \n{(sys)}$ \\
          ${}^{57}\n{Co}$  & 122    & $\gamma$ & $23.60 \pm 0.03 \, \n{(stat)} \pm 0.85 \, \n{(sys)}$\\
          ${}^{83m}\n{Kr}$ & 32.1   & $e^-$    & $27.38 \pm 0.12 \, \n{(stat)} \pm 0.82 \, \n{(sys)}$\\
          ${}^{83m}\n{Kr}$ & 9.4    & $e^-$    & $28.80 \pm 0.08 \, \n{(stat)} \pm 0.86 \, \n{(sys)}$\footnote{This
		  value depends on the time difference between the 32.1 keV and 9.4 keV transitions, see Sec.~\ref{sec:discussion} for details.} \\
          \hline \hline
	\end{tabular*}
\end{table}   

\subsection{Internal ${}^{83m}\n{Kr}$ Irradiation}
\label{subsec:kr83m}

The ${}^{83m}\n{Kr}$ isomer, produced in the decay of ${}^{83}\n{Rb}$
via pure electron capture, decays to the ground state through two
subsequent transitions of 32.1~keV and 9.4~keV, with half-lives of
1.83~h and 154~ns, respectively. Table~\ref{tab:kr83m_decay_scheme}
lists the possible de-excitation channels and their branching ratios
for the two transitions, as well as the different energies of the
electrons emitted in each channel.  Branching ratios were obtained
from theoretical internal conversion coefficients calculated by the
\texttt{BrIcc} program~\cite{Kibedi:2008aa} and fluorescence yields
from Ref.~\onlinecite{Hubbell:1994aa}.  In both cases, most of the
time the energy is carried by internal conversion and Auger electrons.

\begin{table}[htbp]
  \caption{De-excitation channels and branching ratios of the 32.1~keV and 9.4~keV transitions of
    ${}^{83m}\n{Kr}$. For both transitions, most of the time the energy is carried by internal conversion
    electrons (CE) and Auger electrons (A) instead of $\gamma$ rays. Numbers in parentheses correspond to
    electron energies in keV.}
	\label{tab:kr83m_decay_scheme}
 	\begin{tabular*}{\columnwidth}{@{\extracolsep{\fill}} c l c } 
		\hline \hline
		Transition & Decay Mode & Branching \\
		Energy &  & Ratio [\%]\\                                                                    
		\hline
		32.1 keV  &$\n{CE}_{M,N}(32)$  & 11.5 \\
                & $\n{CE}_{L}(30.4)+\n{A}(1.6)$  & 63.8 \\
		& $\n{CE}_{K}(17.8)+\n{X}_{K\alpha}(12.6)+ \n{A}(1.6)$  & 15.3 \\
		& $\n{CE}_{K}(17.8)+\n{A}(10.8)+2 \n{A}(1.6)$  & 9.4 \\
		& $\gamma$ & $<0.1$ \\
		\hline
		9.4 keV  & $\n{CE}_{L}(7.5) + \n{A}(1.6)$ & 81.1 \\
		& $\n{CE}_{M}(9.1)$ & 13.1 \\
		& $\gamma$ & 5.8 \\
		\hline \hline
	\end{tabular*}
 \end{table}

 The use of ${}^{83m}\n{Kr}$ as a calibration source allows a uniform
 internal irradiation of the LXe detector, eliminating most of the
 problems mentioned earlier concerning low-energy calibrations with
 external $\gamma$-ray sources. Additionally, the scintillation
 signals produced in LXe by the two subsequent ${}^{83m}\n{Kr}$
 transitions can be separated in time and thus provide precise
 scintillation yield measurements with negligible background
 contribution~\cite{Manalaysay:2009yq}, even at low source activities.
 Since the bulk of the energy in the 32.1~keV transition of
 ${}^{83m}\n{Kr}$ is most often carried by a 30.4~keV conversion
 electron, its scintillation response should provide an independent
 verification of the scintillation yield at that energy obtained in
 the Compton coincidence measurement. Similarly, the scintillation
 response of the 9.4~keV transition is expected to be similar to that
 obtained in the Compton coincidence measurement.

The source used for the irradiation was composed of zeolite beads
containing ${}^{83}\n{Rb}$, which emanate ${}^{83m}\n{Kr}$ from
${}^{83}\n{Rb}$ decays. The ${}^{83}\n{Rb}$ activity of the source
used was $3.45 \1{kBq}$. The source was located in a stainless steel
cylinder connected to the gas system through a $2 \1{\mu m}$ filter
and isolated with a valve. The rate of ${}^{83m}\n{Kr}$ decays
observed was $8 \1{mHz}$, much lower than the activity of the
source. This large reduction in observed rate is attributed to a low
efficiency in the convective transport of ${}^{83m}\n{Kr}$ atoms into
the active volume of the LXe detector. The bulk motion of LXe itself
in and out of the active volume is limited by the small open area
between PMTs and the PTFE holding structure
(Sec.~\ref{sec:setup}). Nevertheless, the distinctive signature of the
two ${}^{83m}\n{Kr}$ transitions allows a clear selection of these
events above background. The measured half-life between the two
transitions is $154 \pm 6 \1{ns}$, in agreement with previously
measured values~\cite{Ahmad:1995zz,Ruby:1963aa}.

\begin{figure}[!htb]
	\begin{center}
		\includegraphics[width=1.0\columnwidth]{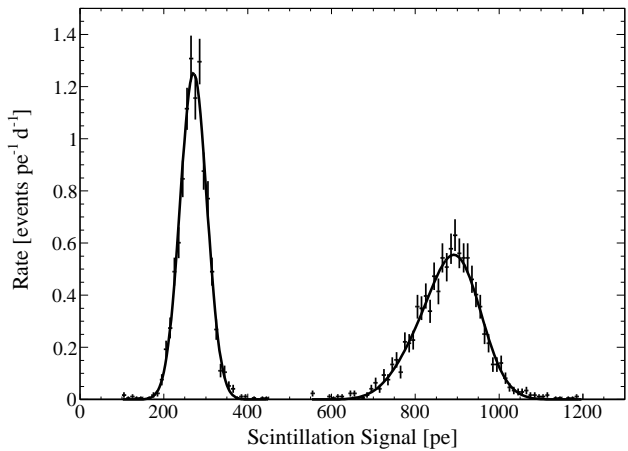}
	\end{center}
	\caption{Measured scintillation spectra (points) for the
          9.4~keV and 32.1~keV de-excitation transitions of
          ${}^{83m}\n{Kr}$, along with their fits (lines). The
          asymmetry of the scintillation spectrum of the 32.1 keV
          transition can be explained by a decrease in the response of
          LXe with decreasing energies.}
	\label{fig:kr83m_s1}
\end{figure}

Fig.~\ref{fig:kr83m_s1} shows the measured scintillation spectra for
the 9.4~keV and 32.1~keV transitions. The scintillation response for
the 9.4~keV transition is compatible with a Gaussian whereas the
response for the 32.1~keV is not and shows a longer tail at low
scintillation values.  The 32.1~keV transition is expected, in about
25\% of cases, to undergo internal conversion with a K-shell electron,
and thus emit a larger number of lower energy electrons than in the
case of internal conversion with an L-shell electron (see
Table~\ref{tab:kr83m_decay_scheme}).  If the scintillation yield of
electrons were to vary with energy then the response of the $32.1
\1{keV}$ transition could have two components.  Therefore, the
response of the 32.1~keV transition is taken as the mean of two
Gaussian functions constrained to have the appropriate branching
ratios. The scintillation light yield value obtained is $27.38 \pm
0.12 \, \n{(stat)} \pm 0.82 \, \n{(sys)} \1{pe/keV}$, with a
resolution $\left({\sigma/E}\right)$ of 6.9\%. The scintillation light
yield of the 9.4~keV transition obtained is $28.80 \pm 0.08 \,
\n{(stat)} \pm 0.86 \, \n{(sys)} \1{pe/keV}$, with a resolution
$\left({\sigma/E}\right)$ of 11.8\%. The measured variation in the PMT
gains during the internal irradiation with ${}^{83m}\n{Kr}$ is taken
as the systematic uncertainty in the light yield. The ratio of the
measured scintillation light yields of the 32.1~keV and 9.4~keV decays
is $1.052 \pm 0.005$, a value consistent with the results of
Ref.~\onlinecite{Manalaysay:2009yq} which found $1.056 \pm 0.011$. In
Ref.~\onlinecite{Kastens:2009pa}, the scintillation yields measured
lead to a slightly lower ratio of $0.976 \pm 0.001$.

The measured scintillation light yields from the internal irradiation
with ${}^{83m}\n{Kr}$ are summarized in Table~\ref{tab:source_ly},
along with the results for external $\gamma$-ray sources.

\section{Results}
\label{sec:results}

The precise determination of the absolute scintillation yield requires
the precise knowledge of many properties related to the scintillation
photon detection probability: the detailed geometry of the active LXe
volume, the reflectivity of the materials, the collection efficiency
of the PMTs and their QE (and its possible variation with
temperature). Thus, relative yields are reported. The reference chosen
is the scintillation yield of the $32.1 \1{keV}$ transition of
$^{83m}\n{Kr}$. The use of a low-energy, uniform, internal source as a
reference has major advantages over an external $\gamma$-ray source
such as $^{57}\n{Co}$: the systematic uncertainty on the $^{57}\n{Co}$
scintillation yield (Sec.~\ref{subsec:gamma_ray_sources}) arising from
the highly localized event distribution in LXe can be
eliminated. Additionally, since the reference source is internal, it
readily solves the problem of the small penetration depth of
low-energy $\gamma$ rays in the calibration of the inner volume of
large detectors.

\begin{figure}[htbp]
	\begin{center}
		\includegraphics[width=1.0\columnwidth]{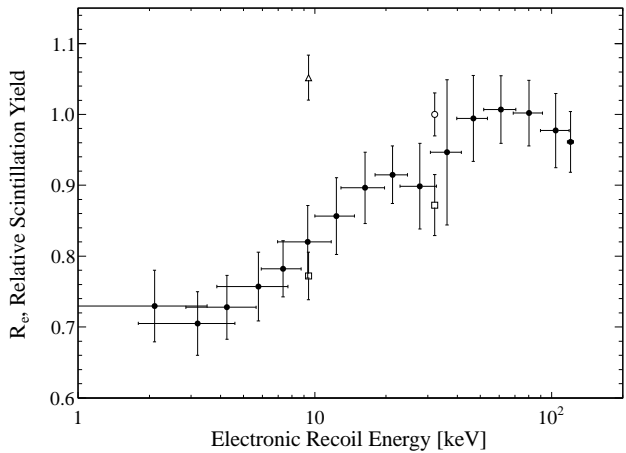}
	\end{center}
	\caption{Measured values (solid circles) of the relative
          scintillation yield of electronic recoils, $\mathcal{R}_e$,
          with respect to the scintillation yield of the $32.1
          \1{keV}$ transition of ${}^{83m}\n{Kr}$ (open circle), along
          with that of the $9.4 \1{keV}$ transition (open triangle).
          The predicted relative yields of the two transitions,
          computed from the Compton coincidence results and the
          electron energies emitted
          (Table~\ref{tab:kr83m_decay_scheme}), are also indicated
          (open squares). The anomalous scintillation yield of the
          $9.4 \1{keV}$ transition of ${}^{83m}\n{Kr}$, compared to
          that of an electronic recoil of the same energy, can be
          understood by the transient state of the LXe after the
          absorption of the electrons emitted in the $32.1 \1{keV}$
          transition, as explained in the text.}
	\label{fig:results}
\end{figure}

The obtained values of the relative scintillation yield of electronic
recoils at zero field, $\mathcal{R}_{e}$, are listed in
Table~\ref{tab:results}. The specific Compton coincidence data sets
used to calculate the $\mathcal{R}_{e}$ values are also listed for
each electronic recoil energy, labelled by the scattering angle
$\theta_\n{HPGe}$ between the $^{137}\n{Cs}$ source and the center of
the LXe and HPGe detectors. Fig.~\ref{fig:results} shows the results
as a function of electronic recoil energy, along with the measured and
predicted relative scintillation yields of the 32.1 and $9.4 \1{keV}$
transitions of $^{83m}\n{Kr}$.

\begin{table}[htbp]
  \caption{Values of the relative scintillation yield of electronic recoils at zero field,
    $\mathcal{R}_{e}$, together with their uncertainties, obtained from different sets of Compton
    coincidence measurements, labelled by the scattering angle $\theta_\n{HPGe}$ between the $^{137}\n{Cs}$
    source and the center of the LXe and HPGe detectors.}
	\label{tab:results}
	\begin{tabular*}{\columnwidth}{@{\extracolsep{\fill}} c l c}
		\hline \hline
		$E_{\n{er}}$ (keV) & Measurements ($\theta_\n{HPGe}$) & $\mathcal{R}_{e}$ \\
		\hline
		$2.1 \pm 1.4$ & 0.0$^\circ$, 5.6$^\circ$ & $0.730 \pm 0.050$ \\
		$3.2 \pm 1.4$ & 0.0$^\circ$, 5.6$^\circ$ & $0.705 \pm 0.045$ \\
                $4.3 \pm 1.4$ & 0.0$^\circ$, 5.6$^\circ$ & $0.728 \pm 0.045$ \\
                $5.8 \pm 1.9$ & 0.0$^\circ$, 5.6$^\circ$, 8.6$^\circ$ & $0.757 \pm 0.048$ \\
		$7.3 \pm 1.4$ & 0.0$^\circ$, 5.6$^\circ$, 8.6$^\circ$ & $0.782 \pm 0.040$ \\
		$9.3 \pm 2.4$ & 0.0$^\circ$, 5.6$^\circ$, 8.6$^\circ$, 12.0$^\circ$ & $0.820 \pm 0.051$ \\
		$12.3 \pm 2.3$ & 0.0$^\circ$, 5.6$^\circ$, 8.6$^\circ$, 12.0$^\circ$ & $0.857 \pm 0.054$ \\
		$16.3 \pm 3.4$ & 0.0$^\circ$, 5.6$^\circ$, 8.6$^\circ$, 12.0$^\circ$ & $0.896 \pm 0.050$ \\
		$21.3 \pm 3.3$ & 0.0$^\circ$, 8.6$^\circ$, 12.0$^\circ$, 16.1$^\circ$ & $0.915 \pm 0.041$ \\
		$27.8 \pm 4.9$ & 0.0$^\circ$, 8.6$^\circ$, 12.0$^\circ$, 16.1$^\circ$ & $0.899 \pm 0.060$ \\
                $36.2 \pm 5.4$ & 16.1$^\circ$, 21.3$^\circ$ & $0.947 \pm 0.103$ \\
		$46.7 \pm 6.9$ & 21.3$^\circ$ & $0.994 \pm 0.061$ \\
		$61.1 \pm 9.4$ & 21.3$^\circ$, 28.1$^\circ$ & $1.007 \pm 0.048$ \\
		$80.2 \pm 11.4$ & 21.3$^\circ$, 28.1$^\circ$, 34.4$^\circ$ & $1.002 \pm 0.046$ \\
                $104.2 \pm 14.4$ & 28.1$^\circ$, 34.4$^\circ$ & $0.977 \pm 0.052$ \\
                $120.2 \pm 3.4$ & 34.4$^\circ$ & $0.961 \pm 0.043$ \\
		\hline \hline
	\end{tabular*}
\end{table}

The statistical uncertainty on $\mathcal{R}_{e}$ comes from the fit of
the electronic recoil peak while the systematic contributions arise
from uncertainties in the PMT gains, $\sigma_{g_i}$, the HPGe
calibration factor, $\sigma_{C_\n{HPGe}}$, and the background
subtraction, $\sigma_b$. The systematic uncertainty arising from the
variation in the fitting range and the spread in electronic recoil
energies were found to have a negligible impact and are therefore not
included. However, the observed variance of $\mathcal{R}_{e}$ values
for the same electronic recoil energy from different measurements was
found to be greater than that given by the contributions mentioned
above. Consequently, an additional term, $\sigma^2_{\mathcal{R}_{e},
  \n{s}}$, is included in the expression for the total uncertainty on
$\mathcal{R}_{e}$ to account for this. The total uncertainty on
$\mathcal{R}_{e}$ is therefore expressed as

\vspace{-10pt}
\begin{multline}
	\sigma^2_{\mathcal{R}_{e}} = \sigma^2_{\mathcal{R}_{e}, \n{fit}}
	+ \sum_i \left({\frac{\partial \mathcal{R}_{e}}{\partial g_i}}\right)^2 \sigma^2_{g_i}\\
	+ \left({\frac{\Delta \mathcal{R}_{e}}{\Delta C_\n{HPGe}}}\right)^2 \sigma^2_{C_\n{HPGe}}
	+ \left({\frac{\Delta \mathcal{R}_{e}}{\Delta b}}\right)^2 \sigma^2_{b}
	+ \sigma^2_{\mathcal{R}_{e}, \n{s}} \n{.}
\end{multline}

The uncertainty in PMT gains is taken as the variation in the measured
gains during the data taking period. The variation in
$\mathcal{R}_{e}$ values with respect to the HPGe detector calibration
was calculated through a finite difference approximation, $\Delta
\mathcal{R}_{e}/\Delta C_\n{HPGe}$, by repeating the analysis using
the calibration factors $C_\n{HPGe} \pm \sigma_{C_\n{HPGe}}$. For
electronic recoil energies below $5 \1{keV}$ ($\theta_\n{HPGe} =$
$0^\circ$, $5.6^\circ$), the contribution from the uncertainty in the
background subtraction was estimated by repeating the analysis with a
different background model.  Specifically, since low-energy background
events are expected to arise from accidental coincidences between LXe
and HPGe detector triggers, as explained in
Sec.~\ref{subsec:extracting_ly}, an alternate background model based
on the energy spectrum of accidental coincidence events was used. LXe
scintillation signal and HPGe energy random variates, distributed
according to the measured LXe and HPGe detector $^{137}\n{Cs}$
spectra, were used to generate the expected background from accidental
coincident events. The background contamination was calculated such
that the resulting LXe scintillation signal spectrum, with the
background spectrum subtracted, matched the rate obtained from the
GEANT4 Monte Carlo simulation. A recoil energy region virtually free
of background, from $10 \1{keV}$ to $20 \1{keV}$, was used to
normalize the simulated rate.

The additional uncertainty contribution $\sigma^2_{\mathcal{R}_{e},
  \n{s}}$ is taken as a linear function of the recoil energy, from
7.1\% at $2 \1{keV}$ down to 3\% at $53 \1{keV}$, and vanishing for
recoil energies above $78 \1{keV}$. For electronic recoil energies
below $53 \1{keV}$, the largest contribution to the uncertainty comes
from this additional contribution. The next largest contribution to
the uncertainty at these energies comes from the uncertainty in the
PMT gains (3\%), which is the same for all measurements. At $2
\1{keV}$, the contribution from the statistical uncertainty (2.8\%),
and those of the background subtraction (0.8\%) and HPGe detector
calibration (0.6\%) are next in size. At recoil energies above $53
\1{keV}$, the contribution from the PMT gains dominates while the
contributions from other effects are negligible.

When multiple $\mathcal{R}_{e}$ values were obtained at a given
electronic recoil energy from different Compton coincidence data sets,
the results were averaged taking the total uncertainty of each value
into account. Similarly, values of $\mathcal{R}_{e}$, which were
calculated at $1 \1{keV}$ HPGe energy intervals, were averaged over
ranges of electronic recoil energies where $\mathcal{R}_{e}$ did not
vary appreciably.

\section{Discussion}
\label{sec:discussion}

To our knowledge, these results are the first measurements of the
scintillation response of LXe to nearly monoenergetic low-energy
electrons over a wide range of energies. The Compton coincidence
technique allows the production of electronic recoils which most
closely resemble the background of large LXe dark matter detectors,
without the need to deconvolve the response for any atomic shell
effects present in the case of the response to low-energy
photo-absorbed $\gamma$ rays.

Our results suggest that the scintillation yield of electronic recoils
at zero field increases as the electron energy decreases from $120
\1{keV}$ to about $60 \1{keV}$ but then decreases by about 30\% from
$60 \1{keV}$ to $2 \1{keV}$, contrary to the intuition that it should
continue to increase with ionization density. This odd behavior is
expected, however, since the electron-ion recombination probability
has been shown to become independent of ionization density for
low-energy electronic recoils~\cite{Dahl:2009th}. For an electronic
recoil track size smaller than the electron thermalization length in
LXe, an increase in ionization density is not accompanied by an
increase in recombination probability as ionization electrons
thermalize in a volume larger than that of the track. In fact, the
energy at which the turnover is observed in our measurement
corresponds very closely to the energy at which the average electronic
recoil track size calculated in Ref.~\onlinecite{Dahl:2009th} reaches
$4.6 \1{\mu m}$, the estimated value for the electron thermalization
length in LXe~\cite{Mozumder:1995aa}. At zero field, these electrons
either recombine at much longer time
scales~\cite{Kubota:1978aa,Kubota:1979aa,Doke:1988aa}, attach to
impurities, or eventually leave the active volume of the detector, in
all cases contributing to the reduction in scintillation light from
recombination.

The scintillation yield obtained from the Compton coincidence
measurement is compatible with the measured yield of the $32.1
\1{keV}$ transition of $^{83m}\n{Kr}$, in which the bulk of the energy
released, as described in Sec.~\ref{subsec:kr83m}, is most often
(75\%) carried by a $30 \1{keV}$ internal conversion electron. The
scintillation yield of the $9.4 \1{keV}$ transition of $^{83m}\n{Kr}$,
however, is not compatible with the value from the Compton coincidence
measurement. Assuredly, such a marked disagreement between the two
measured values prompted a search for possible unaccounted systematic
effects in one or both measurements. A notable difference between an
energetic electron produced in the LXe detector by a $\gamma$-ray
Compton scatter and a conversion electron from the $9.4 \1{keV}$
transition is that the latter is produced a very short time, $ 220
\1{ns}$ on average, after another energetic electron, the $30 \1{keV}$
internal conversion electron from the $32.1 \1{keV}$ transition,
transferred its energy to the LXe. On that time scale, electrons and
positive ions from the track of the $32.1 \1{keV}$ transition
conversion electron which have not recombined might still populate the
immediate vicinity of the Kr atom. In the context of the Thomas-Imel
model~\cite{Thomas:1987zz}, in which recombination depends on the
number of Xe ions, and not on ionization density, the enhancement in
the scintillation yield could be understood as being due to the
effective increase in number of ions left over from the previous
interaction. The fact that the predicted relative yields of the two
transitions (Fig.~\ref{fig:results}, open squares), computed from the
Compton coincidence results and the electron energies emitted
(Table~\ref{tab:kr83m_decay_scheme}) are both lower than the
measurements, is also consistent with the above interpretation. That
is, subsequent de-excitations in the cascade have enhanced
scintillation yields, compared to those of isolated recoiling
electrons of the same energy, since they occur very close in time and
in the immediate vicinity of previous tracks.

\begin{figure}[htbp]
	\begin{center}
		  \includegraphics[width=1.0\columnwidth]{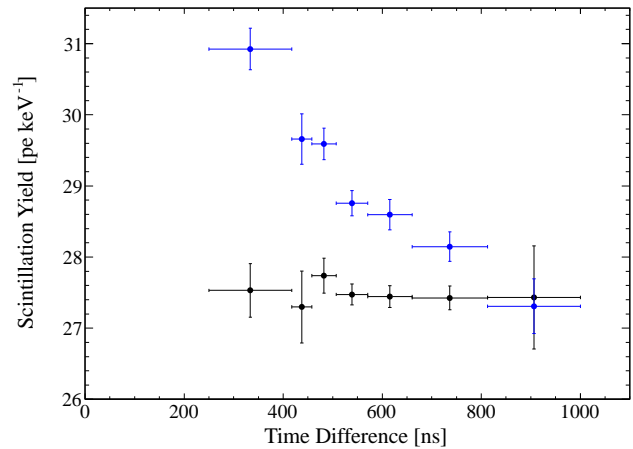}
	\end{center}
	\caption{Scintillation yields of the $9.4 \1{keV}$ (blue) and
          $32.1 \1{keV}$ (black) transitions of ${}^{83m}\n{Kr}$ as
          functions of the time difference between the two
          scintillation signals. While the measured yield of the $32.1
          \1{keV}$ transition is constant with increasing time
          difference, that of the $9.4 \1{keV}$ transition
          decreases. This is a strong indication that the transient
          state of the LXe is responsible for the discrepancy observed
          with respect to the yield measured with the Compton
          coincidence method at this energy.}
	\label{fig:kr83m_9keV_vs_time}
\end{figure}

If the scintillation yield of the $9.4 \1{keV}$ transition of
$^{83m}\n{Kr}$ were to decrease with an increasing time difference
between the two transitions, this would provide a strong indication
that the transient state of the LXe is responsible for the anomalously
high scintillation yield of the $9.4 \1{keV}$ transition, compared to
that measured for Compton electrons of a similar energy.
Fig.~\ref{fig:kr83m_9keV_vs_time} shows the measured scintillation
yields for both transitions as a function of the time difference
between the two scintillation signals. The scintillation yield of the
first transition ($32.1 \1{keV}$) shows no time dependence while that
of the second transition ($9.4 \1{keV}$) exhibits a decrease of $12\%$
from time differences of 300 to $900 \1{ns}$. This raises doubts on
the suitability of $^{83m}\n{Kr}$ as a calibration source in LXe at
$9.4 \1{keV}$, at least at zero electric field.

The efficiency of the data processing software in separating
scintillation signals, which themselves have decay times on the order
of $45 \1{ns}$~\cite{Hitachi:1983zz} at zero field, from two energy
deposits very close in time, such as the two transitions of
$^{83m}\n{Kr}$, necessarily implies a loss in detection efficiency at
short time differences. This efficiency loss, likely different for
measurements from different groups, coupled to a time-dependent
decrease in the scintillation yield of the $9.4 \1{keV}$ transition,
could explain the discrepancy between the ratio of scintillation
yields of the $9.4 \1{keV}$ and $32.1 \1{keV}$ transition of
$^{83m}\n{Kr}$ of this work and the one in
Ref.~\onlinecite{Kastens:2009pa}, a quantity which one would otherwise
expect to be virtually free of most systematic effects.

We have chosen to report relative instead of absolute yields to
eliminate systematic uncertainties in the total light detection
efficiency of the LXe detector from the measurement. The precise
reason for the very high absolute light yield obtained is not known,
although two very likely effects are a temperature dependence of the
QE of the PMTs for LXe scintillation light~\cite{Aprile:2012dy}, and a
change in the effective QE of the PMTs as a function of the angle of
an incident photon~\cite{hamamatsu:pmt_handbook}, the latter being a
much more pronounced effect for a compact detector such as the one
used in this measurement. We have chosen to report our results
relative to the scintillation yield of the $32.1 \1{keV}$ transition
of $^{83m}\n{Kr}$ to minimize the uncertainty from any position
dependence in the light detection efficiency.

We have shown that the improved Compton coincidence
technique~\cite{Choong:2008aa}, with a high energy resolution HPGe
detector, can be used to provide a source of electronic recoils with a
precise energy and small energy spread ($\sim$$1 \1{keV}$). This
technique allows the measurement of the response of LXe to electrons
with energies as low as a few keV, and is only limited by the
resolution of the HPGe detector near the Compton scattered
$\gamma$-ray energy.

\begin{acknowledgments}
  This work was carried out with support from the National Science
  Foundation for the XENON100 Dark Matter experiment at Columbia
  University (Award No. PHYS09-04220).
\end{acknowledgments}



\begin{thebibliography}{33}
\expandafter\ifx\csname natexlab\endcsname\relax\def\natexlab#1{#1}\fi
\expandafter\ifx\csname bibnamefont\endcsname\relax
  \def\bibnamefont#1{#1}\fi
\expandafter\ifx\csname bibfnamefont\endcsname\relax
  \def\bibfnamefont#1{#1}\fi
\expandafter\ifx\csname citenamefont\endcsname\relax
  \def\citenamefont#1{#1}\fi
\expandafter\ifx\csname url\endcsname\relax
  \def\url#1{\texttt{#1}}\fi
\expandafter\ifx\csname urlprefix\endcsname\relax\def\urlprefix{URL }\fi
\providecommand{\bibinfo}[2]{#2}
\providecommand{\eprint}[2][]{\url{#2}}

\bibitem[{\citenamefont{Aprile et~al.}(2010)}]{Aprile:2010um}
\bibinfo{author}{\bibfnamefont{E.}~\bibnamefont{Aprile}} \bibnamefont{et~al.}
  (\bibinfo{collaboration}{XENON100}), \bibinfo{journal}{Phys. Rev. Lett.}
  \textbf{\bibinfo{volume}{105}}, \bibinfo{pages}{131302}
  (\bibinfo{year}{2010}), \eprint{1005.0380}.

\bibitem[{\citenamefont{Aprile et~al.}(2011)}]{Aprile:2011hi}
\bibinfo{author}{\bibfnamefont{E.}~\bibnamefont{Aprile}} \bibnamefont{et~al.}
  (\bibinfo{collaboration}{XENON100}), \bibinfo{journal}{Phys.Rev.Lett.}
  \textbf{\bibinfo{volume}{107}}, \bibinfo{pages}{131302}
  (\bibinfo{year}{2011}), \eprint{1104.2549}.

\bibitem[{\citenamefont{Aprile et~al.}(2012{\natexlab{a}})}]{Aprile:2012nq}
\bibinfo{author}{\bibfnamefont{E.}~\bibnamefont{Aprile}} \bibnamefont{et~al.}
  (\bibinfo{collaboration}{XENON100}) (\bibinfo{year}{2012}{\natexlab{a}}),
  \eprint{1207.5988}.

\bibitem[{\citenamefont{Aprile et~al.}()}]{Aprile:2012aa}
\bibinfo{author}{\bibfnamefont{E.}~\bibnamefont{Aprile}} \bibnamefont{et~al.}
  (\bibinfo{collaboration}{XENON}), \bibinfo{note}{proceedings of UCLA DM 2012
  (to be published)}.

\bibitem[{\citenamefont{Aprile et~al.}(2012{\natexlab{b}})}]{Aprile:2011dd}
\bibinfo{author}{\bibfnamefont{E.}~\bibnamefont{Aprile}} \bibnamefont{et~al.}
  (\bibinfo{collaboration}{XENON100}), \bibinfo{journal}{Astropart. Phys.}
  \textbf{\bibinfo{volume}{35}}, \bibinfo{pages}{573 }
  (\bibinfo{year}{2012}{\natexlab{b}}).

\bibitem[{\citenamefont{Aprile et~al.}(2005)}]{Aprile:2005mt}
\bibinfo{author}{\bibfnamefont{E.}~\bibnamefont{Aprile}} \bibnamefont{et~al.},
  \bibinfo{journal}{Phys. Rev. D} \textbf{\bibinfo{volume}{72}},
  \bibinfo{pages}{072006} (\bibinfo{year}{2005}), \eprint{astro-ph/0503621}.

\bibitem[{\citenamefont{Aprile et~al.}(2009)}]{Aprile:2008rc}
\bibinfo{author}{\bibfnamefont{E.}~\bibnamefont{Aprile}} \bibnamefont{et~al.},
  \bibinfo{journal}{Phys. Rev. C} \textbf{\bibinfo{volume}{79}},
  \bibinfo{pages}{045807} (\bibinfo{year}{2009}), \eprint{0810.0274}.

\bibitem[{\citenamefont{Plante et~al.}(2011)\citenamefont{Plante, Aprile,
  Budnik, Choi, Giboni et~al.}}]{Plante:2011hw}
\bibinfo{author}{\bibfnamefont{G.}~\bibnamefont{Plante}},
  \bibinfo{author}{\bibfnamefont{E.}~\bibnamefont{Aprile}},
  \bibinfo{author}{\bibfnamefont{R.}~\bibnamefont{Budnik}},
  \bibinfo{author}{\bibfnamefont{B.}~\bibnamefont{Choi}},
  \bibinfo{author}{\bibfnamefont{K.}~\bibnamefont{Giboni}},
  \bibnamefont{et~al.}, \bibinfo{journal}{Phys. Rev. C}
  \textbf{\bibinfo{volume}{84}}, \bibinfo{pages}{045805}
  (\bibinfo{year}{2011}), \eprint{1104.2587}.

\bibitem[{\citenamefont{Doke et~al.}(2002)\citenamefont{Doke, Hitachi, Kikuchi,
  Masuda, Okada, and Shibamura}}]{Doke:2002aa}
\bibinfo{author}{\bibfnamefont{T.}~\bibnamefont{Doke}},
  \bibinfo{author}{\bibfnamefont{A.}~\bibnamefont{Hitachi}},
  \bibinfo{author}{\bibfnamefont{J.}~\bibnamefont{Kikuchi}},
  \bibinfo{author}{\bibfnamefont{K.}~\bibnamefont{Masuda}},
  \bibinfo{author}{\bibfnamefont{H.}~\bibnamefont{Okada}}, \bibnamefont{and}
  \bibinfo{author}{\bibfnamefont{E.}~\bibnamefont{Shibamura}},
  \bibinfo{journal}{Jpn. J. Appl. Phys.} \textbf{\bibinfo{volume}{41}},
  \bibinfo{pages}{1538} (\bibinfo{year}{2002}).

\bibitem[{\citenamefont{Doke et~al.}(1988)}]{Doke:1988aa}
\bibinfo{author}{\bibfnamefont{T.}~\bibnamefont{Doke}} \bibnamefont{et~al.},
  \bibinfo{journal}{Nucl. Instrum. Methods Phys. Res., Sect. A}
  \textbf{\bibinfo{volume}{269}}, \bibinfo{pages}{291} (\bibinfo{year}{1988}).

\bibitem[{\citenamefont{Barabanov et~al.}(1987)\citenamefont{Barabanov, Gavrin,
  and Pshukov}}]{Barabanov:1987fj}
\bibinfo{author}{\bibfnamefont{I.~R.} \bibnamefont{Barabanov}},
  \bibinfo{author}{\bibfnamefont{V.~N.} \bibnamefont{Gavrin}},
  \bibnamefont{and} \bibinfo{author}{\bibfnamefont{A.~M.}
  \bibnamefont{Pshukov}}, \bibinfo{journal}{Nucl. Instrum. Meth. Phys. Res.,
  Sect. A} \textbf{\bibinfo{volume}{254}}, \bibinfo{pages}{355}
  (\bibinfo{year}{1987}).

\bibitem[{\citenamefont{Obodovskii and Ospanov}(1994)}]{Obodovskii:1994aa}
\bibinfo{author}{\bibfnamefont{I.~M.} \bibnamefont{Obodovskii}}
  \bibnamefont{and} \bibinfo{author}{\bibfnamefont{K.~T.}
  \bibnamefont{Ospanov}}, \bibinfo{journal}{Instrum. Exp. Tech.}
  \textbf{\bibinfo{volume}{37}}, \bibinfo{pages}{42} (\bibinfo{year}{1994}).

\bibitem[{\citenamefont{Yamashita et~al.}(2004)\citenamefont{Yamashita, Doke,
  Kawasaki, Kikuchi, and Suzuki}}]{Yamashita:2004}
\bibinfo{author}{\bibfnamefont{M.}~\bibnamefont{Yamashita}},
  \bibinfo{author}{\bibfnamefont{T.}~\bibnamefont{Doke}},
  \bibinfo{author}{\bibfnamefont{K.}~\bibnamefont{Kawasaki}},
  \bibinfo{author}{\bibfnamefont{J.}~\bibnamefont{Kikuchi}}, \bibnamefont{and}
  \bibinfo{author}{\bibfnamefont{S.}~\bibnamefont{Suzuki}},
  \bibinfo{journal}{Nucl. Instrum. Meth. Phys. Res., Sect. A}
  \textbf{\bibinfo{volume}{535}}, \bibinfo{pages}{692} (\bibinfo{year}{2004}).

\bibitem[{\citenamefont{Manalaysay et~al.}(2010)}]{Manalaysay:2009yq}
\bibinfo{author}{\bibfnamefont{A.}~\bibnamefont{Manalaysay}}
  \bibnamefont{et~al.}, \bibinfo{journal}{Rev. Sci. Instrum.}
  \textbf{\bibinfo{volume}{81}}, \bibinfo{pages}{073303}
  (\bibinfo{year}{2010}), \eprint{0908.0616}.

\bibitem[{\citenamefont{Kastens et~al.}(2009)\citenamefont{Kastens, Cahn,
  Manzur, and McKinsey}}]{Kastens:2009pa}
\bibinfo{author}{\bibfnamefont{L.}~\bibnamefont{Kastens}},
  \bibinfo{author}{\bibfnamefont{S.}~\bibnamefont{Cahn}},
  \bibinfo{author}{\bibfnamefont{A.}~\bibnamefont{Manzur}}, \bibnamefont{and}
  \bibinfo{author}{\bibfnamefont{D.}~\bibnamefont{McKinsey}},
  \bibinfo{journal}{Phys. Rev. C} \textbf{\bibinfo{volume}{80}},
  \bibinfo{pages}{045809} (\bibinfo{year}{2009}), \eprint{0905.1766}.

\bibitem[{\citenamefont{Valentine and Rooney}(1994)}]{Valentine:1994cs}
\bibinfo{author}{\bibfnamefont{J.}~\bibnamefont{Valentine}} \bibnamefont{and}
  \bibinfo{author}{\bibfnamefont{B.}~\bibnamefont{Rooney}},
  \bibinfo{journal}{Nucl. Instrum. Meth. Phys. Res., Sect. A}
  \textbf{\bibinfo{volume}{353}}, \bibinfo{pages}{37} (\bibinfo{year}{1994}).

\bibitem[{\citenamefont{Rooney and Valentine}(1996)}]{Rooney:1996aa}
\bibinfo{author}{\bibfnamefont{B.}~\bibnamefont{Rooney}} \bibnamefont{and}
  \bibinfo{author}{\bibfnamefont{J.}~\bibnamefont{Valentine}},
  \bibinfo{journal}{Nuclear Science, IEEE Transactions on}
  \textbf{\bibinfo{volume}{43}}, \bibinfo{pages}{1271} (\bibinfo{year}{1996}).

\bibitem[{\citenamefont{Choong et~al.}(2008)\citenamefont{Choong, Vetter,
  Moses, Hull, Payne, Cherepy, and Valentine}}]{Choong:2008aa}
\bibinfo{author}{\bibfnamefont{W.-S.} \bibnamefont{Choong}},
  \bibinfo{author}{\bibfnamefont{K.~M.} \bibnamefont{Vetter}},
  \bibinfo{author}{\bibfnamefont{W.~W.} \bibnamefont{Moses}},
  \bibinfo{author}{\bibfnamefont{G.}~\bibnamefont{Hull}},
  \bibinfo{author}{\bibfnamefont{S.~A.} \bibnamefont{Payne}},
  \bibinfo{author}{\bibfnamefont{N.~J.} \bibnamefont{Cherepy}},
  \bibnamefont{and} \bibinfo{author}{\bibfnamefont{J.~D.}
  \bibnamefont{Valentine}}, \bibinfo{journal}{IEEE Trans. Nucl. Sci.}
  \textbf{\bibinfo{volume}{55}}, \bibinfo{pages}{1753} (\bibinfo{year}{2008}).

\bibitem[{\citenamefont{Jortner et~al.}(1965)\citenamefont{Jortner, Meyer,
  Rice, and Wilson}}]{Jortner:1965aa}
\bibinfo{author}{\bibfnamefont{J.}~\bibnamefont{Jortner}},
  \bibinfo{author}{\bibfnamefont{L.}~\bibnamefont{Meyer}},
  \bibinfo{author}{\bibfnamefont{S.~A.} \bibnamefont{Rice}}, \bibnamefont{and}
  \bibinfo{author}{\bibfnamefont{E.~G.} \bibnamefont{Wilson}},
  \bibinfo{journal}{J. of Phys. D: Appl. Phys.} \textbf{\bibinfo{volume}{28}},
  \bibinfo{pages}{178} (\bibinfo{year}{1965}).

\bibitem[{\citenamefont{Giboni et~al.}(2011)}]{Giboni:2011wx}
\bibinfo{author}{\bibfnamefont{K.~L.} \bibnamefont{Giboni}}
  \bibnamefont{et~al.}, \bibinfo{journal}{JINST} \textbf{\bibinfo{volume}{6}},
  \bibinfo{pages}{03002} (\bibinfo{year}{2011}), \eprint{1103.0986}.

\bibitem[{\citenamefont{Agostinelli et~al.}(2003)}]{Agostinelli:2002hh}
\bibinfo{author}{\bibfnamefont{S.}~\bibnamefont{Agostinelli}}
  \bibnamefont{et~al.} (\bibinfo{collaboration}{GEANT4}),
  \bibinfo{journal}{Nucl. Instrum. Meth. Phys. Res., Sect. A}
  \textbf{\bibinfo{volume}{506}}, \bibinfo{pages}{250} (\bibinfo{year}{2003}).

\bibitem[{\citenamefont{{Hamamatsu Photonics
  K.K.}}(2006)}]{hamamatsu:pmt_handbook}
\bibinfo{author}{\bibnamefont{{Hamamatsu Photonics K.K.}}},
  \emph{\bibinfo{title}{{Photomultiplier Tubes: Basics and Applications}}},
  \bibinfo{organization}{{Hamamatsu Photonics K.K.}}, \bibinfo{edition}{3rd}
  ed. (\bibinfo{year}{2006}),
  \urlprefix\url{http://sales.hamamatsu.com/assets/applications/ETD/pmt_handbook_complete.pdf}.

\bibitem[{\citenamefont{Kibedi et~al.}(2008)\citenamefont{Kibedi, Burrows,
  Trzhaskovskayac, Davidsona, and Nestor~Jr.}}]{Kibedi:2008aa}
\bibinfo{author}{\bibfnamefont{T.}~\bibnamefont{Kibedi}},
  \bibinfo{author}{\bibfnamefont{T.~W.} \bibnamefont{Burrows}},
  \bibinfo{author}{\bibfnamefont{M.~B.} \bibnamefont{Trzhaskovskayac}},
  \bibinfo{author}{\bibfnamefont{P.~M.} \bibnamefont{Davidsona}},
  \bibnamefont{and} \bibinfo{author}{\bibfnamefont{C.~W.}
  \bibnamefont{Nestor~Jr.}}, \bibinfo{journal}{Nucl. Instrum. Meth. Phys. Res.,
  Sect. A} \textbf{\bibinfo{volume}{589}}, \bibinfo{pages}{202}
  (\bibinfo{year}{2008}).

\bibitem[{\citenamefont{Hubbell et~al.}(1994)\citenamefont{Hubbell, Trehan,
  Singh, Chand, Mehta, Garg, Garg, Singh, and Puri}}]{Hubbell:1994aa}
\bibinfo{author}{\bibfnamefont{J.~H.} \bibnamefont{Hubbell}},
  \bibinfo{author}{\bibfnamefont{P.~N.} \bibnamefont{Trehan}},
  \bibinfo{author}{\bibfnamefont{N.}~\bibnamefont{Singh}},
  \bibinfo{author}{\bibfnamefont{B.}~\bibnamefont{Chand}},
  \bibinfo{author}{\bibfnamefont{D.}~\bibnamefont{Mehta}},
  \bibinfo{author}{\bibfnamefont{M.~L.} \bibnamefont{Garg}},
  \bibinfo{author}{\bibfnamefont{R.~R.} \bibnamefont{Garg}},
  \bibinfo{author}{\bibfnamefont{S.}~\bibnamefont{Singh}}, \bibnamefont{and}
  \bibinfo{author}{\bibfnamefont{S.}~\bibnamefont{Puri}}, \bibinfo{journal}{J.
  Phys. Chem. Ref. Data} \textbf{\bibinfo{volume}{23}}, \bibinfo{pages}{339}
  (\bibinfo{year}{1994}).

\bibitem[{\citenamefont{Ahmad et~al.}(1995)\citenamefont{Ahmad, Rehm, Kanter,
  Kutschera, Phillips et~al.}}]{Ahmad:1995zz}
\bibinfo{author}{\bibfnamefont{I.}~\bibnamefont{Ahmad}},
  \bibinfo{author}{\bibfnamefont{K.}~\bibnamefont{Rehm}},
  \bibinfo{author}{\bibfnamefont{E.}~\bibnamefont{Kanter}},
  \bibinfo{author}{\bibfnamefont{W.}~\bibnamefont{Kutschera}},
  \bibinfo{author}{\bibfnamefont{W.}~\bibnamefont{Phillips}},
  \bibnamefont{et~al.}, \bibinfo{journal}{Phys. Rev. C}
  \textbf{\bibinfo{volume}{52}}, \bibinfo{pages}{2240} (\bibinfo{year}{1995}).

\bibitem[{\citenamefont{Ruby et~al.}(1963)\citenamefont{Ruby, Hazoni, and
  Pasternak}}]{Ruby:1963aa}
\bibinfo{author}{\bibfnamefont{S.~L.} \bibnamefont{Ruby}},
  \bibinfo{author}{\bibfnamefont{Y.}~\bibnamefont{Hazoni}}, \bibnamefont{and}
  \bibinfo{author}{\bibfnamefont{M.}~\bibnamefont{Pasternak}},
  \bibinfo{journal}{Phys. Rev.} \textbf{\bibinfo{volume}{129}},
  \bibinfo{pages}{826} (\bibinfo{year}{1963}).

\bibitem[{\citenamefont{Dahl}(2009)}]{Dahl:2009th}
\bibinfo{author}{\bibfnamefont{C.~E.} \bibnamefont{Dahl}}, Ph.D. thesis,
  \bibinfo{school}{Princeton University} (\bibinfo{year}{2009}).

\bibitem[{\citenamefont{Mozumder}(1995)}]{Mozumder:1995aa}
\bibinfo{author}{\bibfnamefont{A.}~\bibnamefont{Mozumder}},
  \bibinfo{journal}{Chem. Phys. Lett.} \textbf{\bibinfo{volume}{245}},
  \bibinfo{pages}{359} (\bibinfo{year}{1995}).

\bibitem[{\citenamefont{Kubota et~al.}(1978)}]{Kubota:1978aa}
\bibinfo{author}{\bibfnamefont{S.}~\bibnamefont{Kubota}} \bibnamefont{et~al.},
  \bibinfo{journal}{Phys. Rev. B} \textbf{\bibinfo{volume}{17}},
  \bibinfo{pages}{2762} (\bibinfo{year}{1978}).

\bibitem[{\citenamefont{Kubota et~al.}(1979)\citenamefont{Kubota, Hishida,
  Suzuki, and Ruan(Gen)}}]{Kubota:1979aa}
\bibinfo{author}{\bibfnamefont{S.}~\bibnamefont{Kubota}},
  \bibinfo{author}{\bibfnamefont{M.}~\bibnamefont{Hishida}},
  \bibinfo{author}{\bibfnamefont{M.}~\bibnamefont{Suzuki}}, \bibnamefont{and}
  \bibinfo{author}{\bibfnamefont{J.-z.} \bibnamefont{Ruan(Gen)}},
  \bibinfo{journal}{Phys. Rev. B} \textbf{\bibinfo{volume}{20}},
  \bibinfo{pages}{3486} (\bibinfo{year}{1979}).

\bibitem[{\citenamefont{Thomas and Imel}(1987)}]{Thomas:1987zz}
\bibinfo{author}{\bibfnamefont{J.}~\bibnamefont{Thomas}} \bibnamefont{and}
  \bibinfo{author}{\bibfnamefont{D.}~\bibnamefont{Imel}},
  \bibinfo{journal}{Phys. Rev. A} \textbf{\bibinfo{volume}{36}},
  \bibinfo{pages}{614} (\bibinfo{year}{1987}).

\bibitem[{\citenamefont{Hitachi et~al.}(1983)}]{Hitachi:1983zz}
\bibinfo{author}{\bibfnamefont{A.}~\bibnamefont{Hitachi}} \bibnamefont{et~al.},
  \bibinfo{journal}{Phys.Rev.} \textbf{\bibinfo{volume}{B27}},
  \bibinfo{pages}{5279} (\bibinfo{year}{1983}).

\bibitem[{\citenamefont{Aprile et~al.}(2012{\natexlab{c}})\citenamefont{Aprile,
  Beck, Bokeloh, Budnik, Choi et~al.}}]{Aprile:2012dy}
\bibinfo{author}{\bibfnamefont{E.}~\bibnamefont{Aprile}},
  \bibinfo{author}{\bibfnamefont{M.}~\bibnamefont{Beck}},
  \bibinfo{author}{\bibfnamefont{K.}~\bibnamefont{Bokeloh}},
  \bibinfo{author}{\bibfnamefont{R.}~\bibnamefont{Budnik}},
  \bibinfo{author}{\bibfnamefont{B.}~\bibnamefont{Choi}}, \bibnamefont{et~al.}
  (\bibinfo{year}{2012}{\natexlab{c}}), \eprint{1207.5432}.

\end{thebibliography}

\end{document}